\documentclass[preprint,12pt]{elsarticle}

\usepackage{mathrsfs}
\usepackage{booktabs}
\usepackage{xcolor}
\usepackage{amssymb}
\usepackage{amsmath}
\usepackage{pifont}
\usepackage[colorlinks=true, linkcolor=blue, urlcolor=blue, citecolor=blue]{hyperref}

\usepackage{subcaption}

\begin{document}

\begin{frontmatter}



\title{Energy Conditions and Stability of Charged Wormholes in $f(R, \mathscr{L}_m)$ Gravity: A Comparative Analysis with Compact Objects}


\author[label1]{Sagar V. Soni}
\ead{sagar.soni7878@gmail.com}
\affiliation[label1]{organization={Department of Mathematics, Sardar Patel University},
           city={Vallabh Vidyanagar},
           postcode={388120}, 
           state={Gujarat},
            country={India}}

 \author[label2]{A. C. Khunt}
 \ead{ack.gravity@gmail.com}
\affiliation[label2]{organization={The Gravity Room, Champathal},
            city={Amreli},
            postcode={365601}, 
            state={Gujarat},
            country={India}}          

\author[label3]{Farook Rahaman}
\ead{rahaman@associates.iucaa.in}
\affiliation[label3]{organization={Department of Mathematics, Jadavpur University},
            city={Kolkata},
            postcode={700032}, 
            state={West Bengal},
            country={India}}
                                       
\author[label1]{A. H. Hasmani}
\ead{ah_hasmani@spuvvn.edu}

\begin{abstract}
In this paper, we study the energy conditions of charged traversable wormholes in the framework of $f(R, \mathscr{L}_m)$
 modified gravity. In the first case, we derive the shape functions (SFs) for two different choices of the charge function 
$\mathcal{E}^2$ by considering the Exponential Spheroid (ES) model and analyze the null energy condition (NEC). In the second case, we consider a particular shape function and study its implications for the energy conditions. In both cases, we obtain expressions for energy density and pressure in radial and tangential directions. Our findings show that the radial NEC remains satisfied across a wide range of charge parameters $\mathcal{E}$ consistent with established physical laws. However, the tangential NEC is only sustained in the range $0.1 \leq \mathcal{E} \leq 0.6$; for higher charge values, violations occur, indicating the formation of a throat-like structure necessary for wormhole stability. Additionally, we compare the pressure-density profiles of these charged wormholes with those of compact objects such as neutron stars, revealing distinct variations in matter distribution. This analysis highlights the crucial role of charge and modified gravity in determining the stability and physical characteristics of wormhole structures.
\end{abstract}



\end{frontmatter}



\section{Introduction}
\label{sec1}
In recent years, hypothetical tunnels connecting distant regions of spacetime, known as wormholes, have attracted immense attention due to their profound implications for theoretical physics and cosmology. The exploration of wormholes commenced in 1916 when Flamm \cite{flm} presented the mathematical framework for these entities. In response to the instability observed in earlier solutions, Einstein and Rosen \cite{ER} proposed the concept of the Einstein–Rosen bridge, which provided a theoretical bridge-like connection within the structure of spacetime. Eventually, it was found that the Einstein-Rosen bridge is not traversable. Notably, the effect of electromagnetic interactions in Einstein's field equation may be crucial for the creation of wormholes with geons \cite{geon1, geon2}.  Morris and Thorne \cite{mt} further developed the concept of a traversable wormhole by eliminating the presence of an event horizon. \\
\indent In the classical theory of relativity, wormhole solutions violate all energy conditions \cite{vis95}. To address this, the notion of exotic matter was proposed, enabling theoretical traversability, yet the inquiry into their physical existence persists to be unanswered. The pursuit of overcoming the necessity for exotic matter or relieving energy conditions has driven the investigation of alternate gravitational theories. \\
\indent To avoid reliance on exotic matter, researchers in this area have been attempting a variety of modified theories, notably Einstein-Cartan theory (ECT), teleparallel gravity, $ f(Q)$ theory, $f(R)$ theory of gravitation, and many more. In ECT, researchers \cite{Bron15, soni24} have obtained the solution of wormhole. The traversable wormholes are explored in $f(Q)$ gravity \cite{faruk21, rastgoo, saibal24, symjan} and $f(R)$ gravity \cite{Lobo2009, Bronnikov2007, Pavlovic2015, Mazharimousavi2016, Bahamonde2018, Moraes2017, Godani2019, Capozziello2021}.\\
\indent An expansion of the $f(R)$ theories of gravity, Bertolami et al. \cite{ber2007} investigated the consequences of direct coupling of the scalar curvature $R$ with the matter Lagrangian $\mathscr{L}_m$, derived the mass-particle equation of motion and found an extra force perpendicular to the four-velocity. Based on this structure, Harko and Lobo \cite{hlobo} used it to extend $f(R)$ gravity to $f(R,\mathscr{L}_m)$ theory, where they derived field equations in the metric formalism and demonstrated that test particle motion is typically non-geodesic because of an additional force perpendicular to the four-velocity. Jaybhaye et. al. \cite{lakhan1} have analyzed cosmology within the $f(R,\mathscr{L}_m)$ gravity framework. Researchers have investigated the geometry of wormholes for various models of the $f(R,\mathscr{L}_m)$ \cite{sahoo11, sahoo12, nasir, sahoo13}. Recently, Pawde et al. \cite{mapari} have explored the anisotropic behavior of the universe in $f(R,\mathscr{L}_m)$ gravity.\\
\indent An alternative way to reduce the necessity of exotic matter is to incorporate charge into the framework \cite{mo_17}. Introducing charged fluids as the source for energy-momentum tensor provides an adequate framework for exploring the interactions among gravitational, electromagnetic, and matter fields. The electromagnetic effects that arise from this approach could eliminate the need for exotic matter to fulfill the requirements of spacetime geometry. Charged fluids are significant as they can reduce repulsive gravitational forces and improve the stability of the wormhole structure.The electric charge of the fluid can generate additional forces that prevent energy condition violations, which is a vital aspect of the possibility of traversable wormholes.\\
\indent The inclusion of electric charge into traversable wormhole geometries, as demonstrated by Kim and Lee \cite{kimlee}, represents a notable expansion of the Morris-Thorne wormhole model. This modification stabilizes wormhole geometry using electromagnetic fields, eliminating the need for exotic matter at the throat. Eiroa and Romero \cite{eir2004} studied the linearized stability of charged thin-shell wormholes and found that higher charge values enhance stability by alleviating constraints on the equation of state. Researchers \cite{j09, bro5, j091} examined the feasibility of stabilizing a wormhole in ghost scalar field models and scalar-tensor theories of gravity by incorporating electric charge. In $f(T)$ gravity, charged wormhole were studied by Sharif and Rani \cite{sharif14}. Moraes et al. \cite{mor} derived solutions for charged wormholes in $f(R,T)$-extended gravity, demonstrating that charge enhances stability and permits the fulfillment of energy conditions.\\
\indent The aforementioned literature motivated us to explore traversable wormhole solutions along with a charge in the $f(R, \mathscr{L}_m )$ gravity framework. By selecting a specific model of $f(R, \mathscr{L}_m)$ with the specified Lagrangian $\mathscr{L}_m$, we will investigate the influence of charge on wormholes in two different ways for constant red-shift function. In the first case, we choose energy density using the ES model and with two various choices of $\mathcal{E}$, we generate a SF. For the second case we fix the SF and using the equation of state (EoS) parameter, we find out energy and pressures. We also analyze energy conditions and their stability for both cases.
\section{Theory of $f(R,\mathscr{L}_m)$ Gravitation}
\label{sec2}
In the framework of different theories of gravitation, one of the approaches is generalizing the action as it describes the influence of field equations. In 2010, Harko and Lobo \cite{hlobo} presented the action principle for the $f(R,\mathscr{L}_m)$ gravity model. This action is defined with the help of matter Lagrangian density $\mathscr{L}_m$ and Ricci scalar $R$ and expressed by 
\begin{align}\label{1}
    S=\int f(R,\mathscr{L}_m) \sqrt{-g} d^4x,
\end{align}
where, $f$ is arbitrary function of $R$ and $\mathscr{L}_m$.\\
\indent The Ricci scalar $R$ is obtained by the contraction of Ricci tensor $R_{ij}$, i.e.
\begin{align}
    R=g^{ij}R_{ij},
\end{align}
where, 
\begin{align}
    R_{ij}=\partial_k \Gamma^k_{ij}-\partial_j \Gamma^k_{ki}+\Gamma^h_{ij}\Gamma^k_{hk}-\Gamma^k_{jh}\Gamma^h_{ki},
\end{align}
and $ \Gamma^k_{ij}$ represents the components of the Levi–Civita connection given by
\begin{align}
    \Gamma^k_{ij}=\frac{1}{2} g^{kh}\left(\frac{\partial g_{jh}}{\partial x^i}+\frac{\partial g_{hi}}{\partial x^j}-\frac{\partial g_{ij}}{\partial x^h}\right).
\end{align}
\indent The field equations for $f(R,\mathscr{L}_m)$ gravity can be obtained by taking the variation of action, as provided in equation (\ref{1}), with respect to the metric tensor $g_{ij}$. These equations can be represented as:
\begin{align}\label{EFE}
    f_R R_{ij}-\frac{1}{2}(f-f_{\mathscr{L}_m} \mathscr{L}_m)g_{ij}+(g_{ij}\Box-\nabla_i \nabla_j)f_R=\frac{1}{2} f_{\mathscr{L}_m} T_{ij},
\end{align}
where, $f_R \equiv \frac{\partial f}{\partial R}$, $ f_{\mathscr{L}_m} \equiv \frac{\partial f}{\partial \mathscr{L}_m}$, $\Box \equiv \nabla_i \nabla^i $, and $T_{ij}$ is energy-momentum tensor, which can be expressed as:
\begin{align}
    T_{ij}=\frac{-2}{\sqrt{-g}} \frac{\delta (\sqrt{-g} \mathscr{L}_m)} {\delta g^{ij}}.
\end{align}
\indent Moreover, the contraction of field equations (\ref{EFE}) gives the relation of energy-momentum scalar $T$, matter Lagrangian $\mathscr{L}_m$, and Ricci scalar $R$ as:
\begin{align}
    Rf_R+3\Box f_R-2(f-f_{\mathscr{L}_m }\mathscr{L}_m ) =\frac{1}{2} f_{\mathscr{L}_m} T,
\end{align}
where, $\Box F= \frac{1}{\sqrt{-g
}}\partial_i (\sqrt{-g} g^{ij} \partial_j F)$ for any scalar function $F$.

\section{ Stable traversable wormholes under the Morris-Thorne spacetime}
\label{sec3}
The metric of the Morris-Thorne wormhole, which exhibits spherical symmetry and static behavior, is expressed as follows \cite{mt}:
\begin{align}
    \mathrm{d}s^2=-e^{2{\Phi(r)}}\mathrm{d}t^2+\left(1-\frac{b(r)}{r}\right)^{-1}\mathrm{d}r^2+r^2 \mathrm{d}\theta^2+r^2 \sin^2\theta \mathrm{d}\phi^2,
\end{align}
where, $\Phi(r)$ is red-shift function and $b(r)$ is shape function (SF). This shape function directly affects on the curvature and structure of the throat region. The existence of wormhole solutions requires the satisfaction of the following conditions \cite{mt, vis95}:
\begin{enumerate}[(i)]
  \item $b(r_{0})=r_{0}$, \label{c1}
   \item  For all $r > r_{0}$, $\frac{b(r)-b'(r)r}{b^2(r)}>0$, this is called flare-out condition, \label{c2}
   \item $b'(r)< 1$, \label{c3}
   \item  $\frac{b(r)}{r} < 1$ for $r> r_{0}$, \label{c4}
  \item  $\frac{b(r)}{r}\rightarrow$ as $r\rightarrow \infty$, \label{c5}
  \item $\Phi(r)$ should be finite everywhere.  \label{c6}
 \end{enumerate}
 Also, note that in (\ref{c2}), (\ref{c3}), and (\ref{c4}), the equality conditions hold at the throat.\\
 \indent 
For the unit time-like velocity vector $u_i=(e^{-\Phi(r)},0,0,0)$ and space-like vector $v_i=\left(0,\sqrt{1-\frac{b(r)}{r}},0,0\right)$, we have taken energy-momentum tensor filled with charged anisotropic fluid, expressed as: 
 \begin{align}\label{emt}
     T_{ij}&={(\rho+p_t)u_i u_j +p_t g_{ij}+(p_r-p_t)v_i v_j}\nonumber \\
     & \hspace{0.5cm}+\frac{1}{4\pi}\left[ F_{ik}F^k{ }_j-\frac{1}{4\pi}g_{ij}F_{kh}F^{kh}   \right],
 \end{align}
 where, $\rho$ is the energy density, $p_r$ denotes the radial pressure, and $p_t$ represents the tangential pressure.  The $F_{ij}$ is the skew-symmetric electromagnetic field tensor which is defined in terms of 4-potential $A_i$ by
 \begin{align}
     F_{ij}= A_{j;i}-A_{i;j},
 \end{align}
 and it satisfies the Maxwell equations
 \begin{align}
     F_{ij;k}+ F_{jk;i}+ F_{ki;j}=0,
 \end{align}
 and 
 \begin{align}
   (F^{ik} \sqrt{-g})_{;k}=4\pi \sqrt{-g} J^i,
 \end{align}
 where $J^i$ is the current four vector defined by charge density $\sigma$ as
 \begin{align}
     J^i=\sigma(r) u^i.
 \end{align}
 \indent The only non-vanishing components of $F_{ij}$ is $F_{01}$ and the electric field intensity  $\mathcal{E}$ is obtained from $\mathcal{E}^2=-F_{01}F^{01}$, which reduces to
 \begin{align}
    \mathcal{E}=\frac{Q(r)}{r^2}.
 \end{align}
 where the $Q$ denotes the total charge. We use geometrized units, in which the electric charge $Q$ is dimensionless and expressed in the same units as length or mass, and also \( G = c = 1 \).

 \subsection{Field Equations}
 For the energy-momentum tensor given in (\ref{emt}), the field equations (\ref{EFE}) are written as:
 
 \begin{eqnarray}\label{efe1}
& \left(1 - \frac{b(r)}{r}\right) \left[\left(\Phi'' + \Phi'^2 + \frac{2\Phi}{r} - \frac{rb' - b}{2r(r - b)} \Phi'\right) F \right. \nonumber\\
& \left. - \left(\Phi' + \frac{2}{r} - \frac{rb' - b}{2r(r - b)}\right) F' - F'' \right] \nonumber \\
& + \frac{1}{2} (f - \mathscr{L}_m f_{\mathscr{L}_m}) = \frac{1}{2} f_{\mathscr{L}_m} (\rho +\mathcal{E}^2),
\end{eqnarray}

\begin{eqnarray}\label{efe2}
     &\left(1-\frac{b(r)}{r}\right) \left[\left(-\Phi''-\Phi'^2-\frac{rb'-b}{2r(r-b)}\left(\Phi'+\frac{2}{r}\right)\right)F \right. \nonumber\\
   &\left. +\left(\Phi'+\frac{2}{r}-\frac{rb'-b}{2r(r-b)}\right)F'\right] \nonumber\\
   &+\frac{1}{2}(f-\mathscr{L}_mf_{\mathscr{L}_m})=\frac{1}{2}f_{\mathscr{L}_m}(p_r-\mathcal{E}^2).
\end{eqnarray}

\begin{eqnarray}\label{efe3}
     &\left(1-\frac{b(r)}{r}\right) \left[\left(-\frac{\Phi'}{r}-\frac{rb'+b}{2r^2(r-b)}\right)F \right. \nonumber\\
   &\left.+\left(\Phi'+\frac{2}{r}-\frac{rb'-b}{2r(r-b)}\right)F'-F''\right] \nonumber\\
   &-\frac{1}{2}(f-\mathscr{L}_mf_{\mathscr{L}_m})=\frac{1}{2}f_{\mathscr{L}_m}(p_t+\mathcal{E}^2). 
\end{eqnarray}
 
\section{Solution under specific $f(R,\mathscr{L}_m)$ Model }\label{4}
A general formulation of non-minimal coupling has been presented in the significant work of Harko \cite{h1}. Accordingly, a form of $f(R, \mathscr{L}_m)$ is assumed as $ f(R, \mathscr{L}_m)=f_1(R)+f_2(R)H(\mathscr{L}_m)$, where, $f_1$ and $f_2$ are the function of Ricci scalar $R$ and $H$ is function of matter Lagrangian $\mathscr{L}_m$. To derive solutions for wormholes, we use a specific minimal form of the $f(R,\mathscr{L}_m)$ function, which is provided by \cite{sahoo1, hlobo2}
\begin{align}\label{model}
    f(R, \mathscr{L}_m)=\frac{R}{2}+\mathscr{L}_m^\alpha,
\end{align}
where, $\alpha$ is the free model parameter. The well-known wormhole geometry evolves effortlessly within the classical framework of general relativity by the taking value of $\alpha=1$.\\
\indent The Lagrangian density for the specific choice of the $f(R,\mathscr{L}_m)$ model is defined as the function of energy density and is stated to be $\mathscr{L}_m=\rho$, this choice of $\mathscr{L}_m$ is motivated by the work of Harko and other researchers \cite{h1, l1, hlobo3}. \\
\indent Therefore, for the specific model of $f(R,\mathscr{L}_m)$ given in (\ref{model}) with the chosen Lagrangian $\mathscr{L}_m$, the field equations (\ref{efe1})-(\ref{efe3}) reduce to 

\begin{eqnarray}\label{efe4}
    \frac{b'}{r^2}=\left[(2\alpha-1)\rho+\alpha \mathcal{E}^2\right] \rho^{\alpha-1},
\end{eqnarray}

\begin{eqnarray}\label{efe5}
    \frac{2\Phi'}{r}\left(1-\frac{b}{r}\right)-\frac{b}{r^3}=\left[\alpha p_r-(\alpha-1)\rho-\alpha \mathcal{E}^2\right] \rho^{\alpha-1},
\end{eqnarray}

\begin{eqnarray}\label{efe6}
    &\left[\Phi''+\Phi'^2+\frac{\Phi'}{r}\right]\left(1-\frac{b}{r}\right)-\left[\Phi'+\frac{1}{r}\right]\left(\frac{rb'-b}{2r^2}\right)\nonumber\\
    &=\left[\alpha p_t-(\alpha-1)\rho+\alpha \mathcal{E}^2\right] \rho^{\alpha-1}.
\end{eqnarray}
To solve these field equations, we considered two different cases with a constant red-shift function as follows :
\subsection{\textbf{Case-I}}
\begin{figure}[ht]
    \centering
    \includegraphics[scale=0.5]{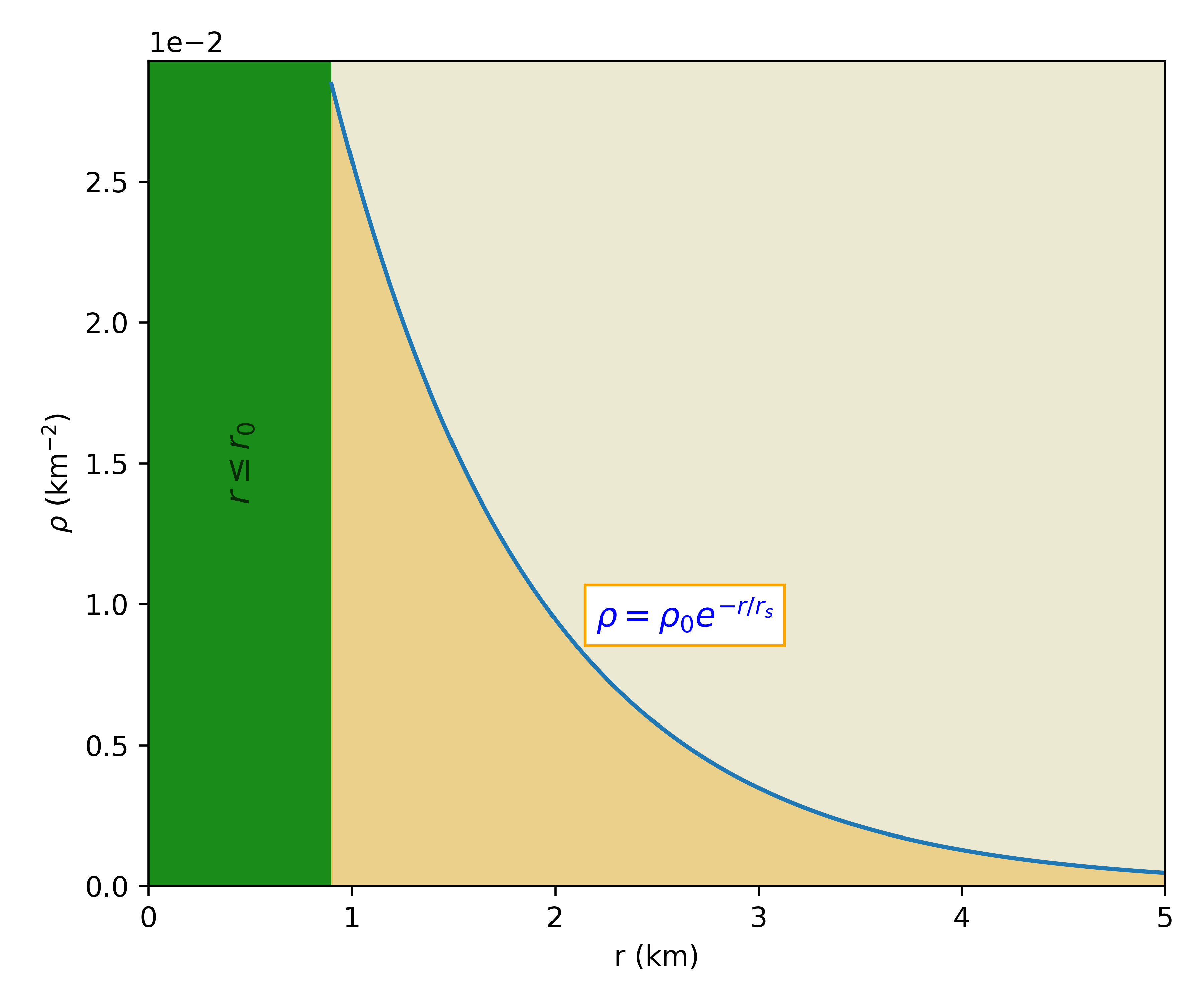}
    \caption{Density $\rho$ plotted against the radial $r$ for the ES-model}
    \label{density}
\end{figure}

This subsection explores the energy density profile related to the ES model. The ES model is described by the following equation \cite{sofue13}
\begin{align}\label{4.1.1}
    \rho=\rho_0 e^{\frac{-r}{r_s}},
\end{align}
where, $\rho_o$, and $r_s$, are free parameters that represent the central dark matter density and the scale radius. Thus, based on equations (\ref{efe4}) and (\ref{4.1.1}), the EC model is governed by the differential equation
\begin{align}\label{4.1.2}
\frac{b'}{r^2}=\left[ (2\alpha-1) \rho_0 e^{\frac{-r}{r_s}}+ \alpha \mathcal{E}^2 \right] \rho_0^{\alpha-1}  e^{\frac{-r (\alpha-1)}{r_s}}.
\end{align}
To solve the aforementioned differential equation, we consider two distinct scenarios for $\mathcal{E}^2$, assuming the value of $r_s=1$. With the help of the expression (\ref{4.1.1}), we shown the density profile in Fig. \ref{density}.

The density profile which was computed using Eq. (\ref{4.1.1}) is shown in Fig. \ref{density}. The density varies gradually from $3\times 10^{-3} $ km$^{-2}$ to $10^{-1}$ km$^{-2}$ over the radial domain extending from $r_{0}$ to $r{\longrightarrow{\infty}}$. This behavior corresponds to the expected physical properties of the system, which include a higher density near $r_{0}$ that decreases as the radius approaches $r_{\infty}$. The profile corresponds to the theoretical predictions for such configurations and allows insight into the spatial distribution of matter within the given framework.

\subsubsection{Shape Function-I:-}
\begin{figure}[ht]
    \centering
    \includegraphics[scale=0.4]{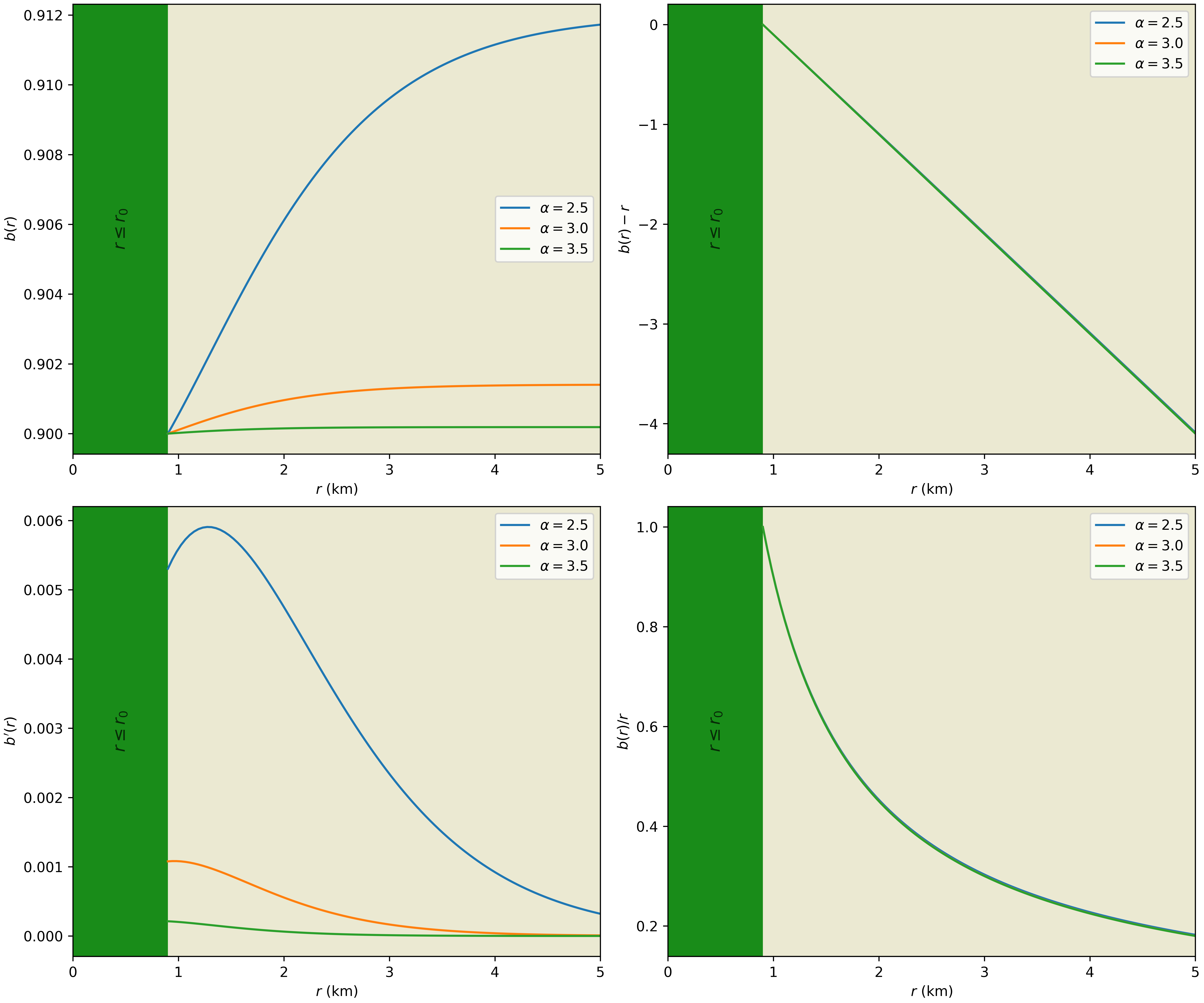}
    \caption{Nature of $b(r)$, throat condition $b(r) < r$, flaring-out condition $b'(r) < 1$, and asymptotic flatness $\lim_{r \to \infty} \frac{b(r)}{r} = 0$ for $r_{0}= 0.9$, $A=0.5$, $\rho_{0}=0.07$, and $\alpha=\{2.5,3.0,3.5\}$.}
    \label{shape1}
\end{figure}

In this case, we choose $\mathcal{E}^2=A$, where $A$ is constant. We derive SF as 
\begin{align}\label{shape1.1}
  b(r) 
    &=
    \frac{\rho_0^{\alpha -1} e^{\alpha  (-r)}}{\alpha^3} \Bigg( 
    -\frac{\alpha ^4 A e^r \left((\alpha -1)^2 r^2+2 (\alpha -1) r+2\right)}{(\alpha -1)^3} \notag \\
    &\quad -(2 \alpha -1) \rho_0 \left(\alpha ^2 r^2+2 \alpha  r+2\right) \Bigg) + C,
\end{align}
where $C$ can be obtained by the condition $b(r_0)=r_0$,
\begin{align}
    C&=r_0+\frac{\rho_0^{\alpha -1} e^{\alpha  (-r_0)}}{\alpha^3} \Bigg( 
    \frac{\alpha ^4 A e^{r_0} \left((\alpha -1)^2 r_0^2+2 (\alpha -1) r_0+2\right)}{(\alpha -1)^3} \notag \\
    &\quad +(2 \alpha -1) \rho_0 \left(\alpha ^2 r_0^2+2 \alpha  r_0+2\right) \Bigg). 
\end{align}
The flaring-out condition is satisfied if the following constraints hold  
\begin{align}\label{9}
\left[ (2\alpha-1) \rho_0^\alpha e^{{-\alpha r_0}}+ \alpha A \rho_0^{\alpha-1}  e^{{-r_0 (\alpha-1)}} \right]<\frac{1}{r_0^2}.
\end{align}
In the case of GR, $\alpha=1$, so by above equation we get following inequality
\begin{align}
    \rho_0<\left[\frac{1}{r_0^2}-A\right]e^{r_0}.
\end{align}
Here, we choose $r_0=0.9$, $A=0.5$ and hence, $\rho_0<1.8$. We have fixed $\rho_0=0.07$ and by inequality (\ref{9}), we get value of $\alpha$ as $\alpha>0.72$.\\
 \indent We will use the graphical analysis to determine the feasibility of the wormhole shape function (\ref{shape1.1}) as well as whether it satisfies the necessary requirements for wormhole existence. Figure \ref{shape1} shows the graphical depiction of $b(r)$, $b(r)-r$, $b'(r)$ and $b(r)/r$. It is crucial to note that for our present investigation, we choose $r_{0}=0.9$, $A=0.5$, and $\rho_{0}=0.07$, in addition to various changes of parameter $\alpha$.
The graphical representation of $b(r)$, which clearly shows that the condition $b(r_{0})=0$ is satisfied at $r_{0}$, where $r_{0}=0.9$ is the the throat of the wormhole. The asymptotically flat necessity condition for the wormhole SF was satisfied, as shown by the function $b(r)/r \rightarrow{0}$ for $r \rightarrow{\infty}$. The plot indicates that $b'(r)>0$ is crucial to formation of the wormhole in its present state.

The radial pressure and tangential pressure can be expressed as 
\begin{align}
  p_r&=  \frac{1}{\alpha}\left[(\alpha-1)\rho_0 e^{-r}+\alpha A+ \frac{ e^{-r}}{\alpha^3 r^3} \Bigg( 
    \frac{\alpha ^4 A e^r \left((\alpha -1)^2 r^2+2 (\alpha -1) r+2\right)}{(\alpha -1)^3}\right. \notag \\
    &\quad \left.+(2 \alpha -1) \rho_0 \left(\alpha ^2 r^2+2 \alpha  r+2\right) \Bigg) \right. \notag \\
    &\quad \left. - \frac{r_0}{r^3} \rho_0^{1-\alpha} e^{(\alpha-1)r}-\frac{ e^{(\alpha-1)r-\alpha  r_0}}{\alpha^3 r^3} \Bigg( 
    \frac{\alpha ^4 A e^{r_0} \left((\alpha -1)^2 r_0^2+2 (\alpha -1) r_0+2\right)}{(\alpha -1)^3} \right. \notag \\
    &\quad\left. +(2 \alpha -1) \rho_0 \left(\alpha ^2 r_0^2+2 \alpha  r_0+2\right) \Bigg) \right],
    \end{align}
    \begin{align}
     p_t&=\frac{1}{\alpha}\left[(\alpha-1)\rho_0 e^{-r}-\alpha A \right.\notag\\
     &\quad \left. +e^{(\alpha -1) r} \rho_0^{1-\alpha } \Bigg( \frac{1}{2 r^3} \Bigg[ \frac{\rho_0^{\alpha -1} e^{\alpha  (-r)} }{\alpha^3} \Bigg( 
    -\frac{\alpha ^4 A e^r \left((\alpha -1)^2 r^2+2 (\alpha -1) r+2\right)}{(\alpha -1)^3} \right. \notag \\
    &\quad \left. -(2 \alpha -1) \rho_0 \left(\alpha ^2 r^2+2 \alpha  r+2\right) \Bigg) 
    + r_0 \right. \notag\\
    &\quad \left. + \frac{\rho_0^{\alpha -1} e^{\alpha  (-r_0)}}{\alpha^3} \Bigg( 
    \frac{\alpha ^4 A e^{r_0} \left((\alpha -1)^2 r_0^2+2 (\alpha -1) r_0+2\right)}{(\alpha -1)^3} \right. \notag \\
    &\quad \left. +(2 \alpha -1) \rho_0 \left(\alpha ^2 r_0^2+2 \alpha  r_0+2\right) \Bigg) \Bigg]  \right. \notag \\
    &\quad \left. -\frac{1}{2 r^2} \Bigg[ r^2 \rho_0^{\alpha -1} e^{(\alpha -1) (-r)} \left(\alpha  A+(2 \alpha -1) \rho_0 e^{-r}\right) \Bigg] \Bigg) \right].
\end{align}
\begin{figure}[ht]
    \centering
    \includegraphics[scale=0.4]{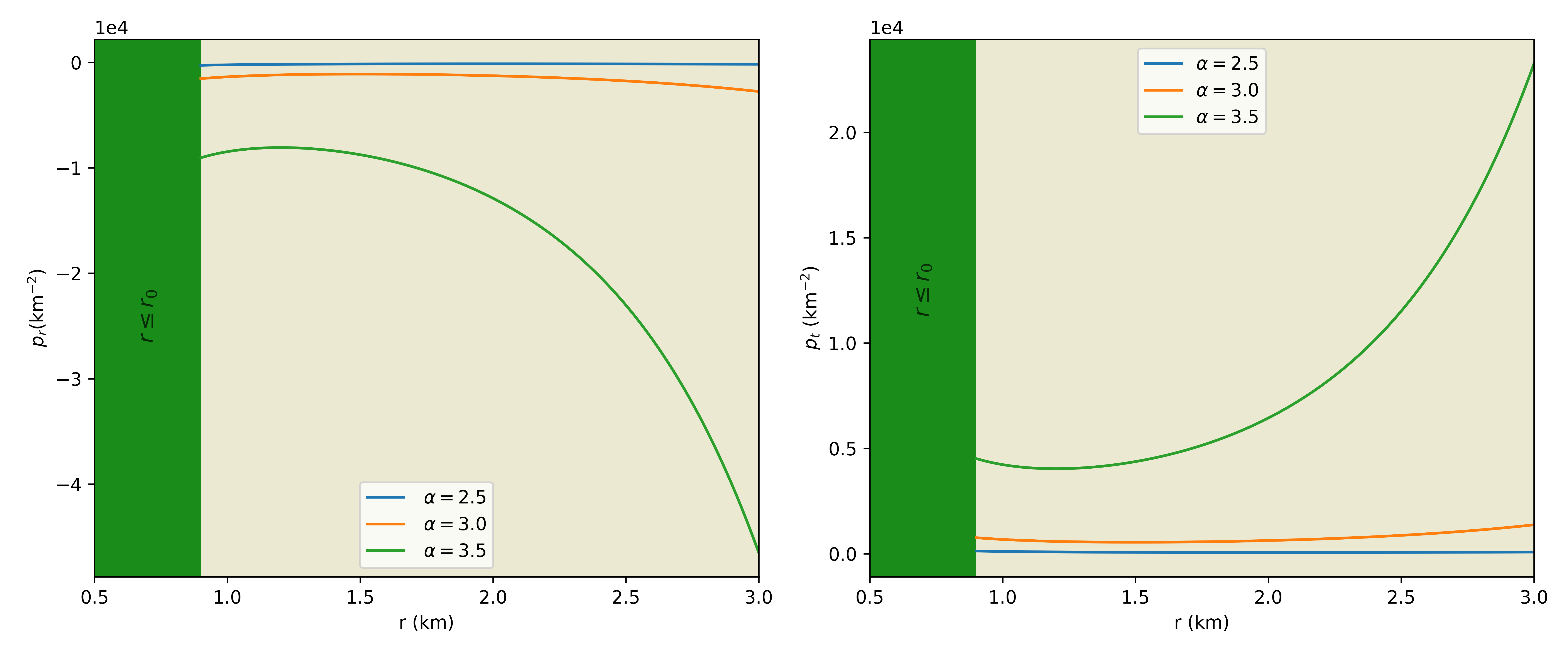}
    \caption{\textit{Left}: Radial pressure $p_r$ and \textit{Right}: Tangential pressure $p_t$ against the radius $r$ for the ES model. The graph shows different curve for the values of $\alpha$ for 2.5, 3.0, and 3.5, with the model parameter set to $A=0.5$, the density parameter $\rho_0=0.07$ and $r_0=0.9$ for the charge function $\mathcal{E}^2=A$.}
    \label{pr_pt_1}
\end{figure}
The graphical illustration of radial pressure and tangential pressure can be seen in Fig. \ref{pr_pt_1}.  It can be noted that radial pressure has negative tendencies whereas tangential pressure is positive and increasing in nature. It is relevant to emphasize here that we choose $r_{0}=0.9$, $A=0.5$, and $\rho_{0}=0.07$, and for different variations of parameter $\alpha$ as mentioned on the graph.

Within the wormhole structure, particularly in the vicinity of the throat, the energy density $\rho$ shows strictly positive values in its graphical behavior. Instead of exotic matter, which generally corresponds to negative energy density, this supports the existence of baryonic matter. An anisotropic nature of the matter distribution is further shown by the tangential pressure, $p_{t}$, being positive while the radial pressure, $p_{r}$, exhibits negative values. Without requiring an escape of energy conditions, the observed pressure anisotropy suggests that the wormhole remains stable by an asymmetric force distribution. We present an anisotropy factor, $\Delta (=p_{t}-p_{r})$, which is shown in Fig. \ref{Delta-1} for different values of $\alpha$.

\subsubsection{Shape Function-II:-}
\begin{figure}[ht]
    \centering
    \includegraphics[scale=0.4]{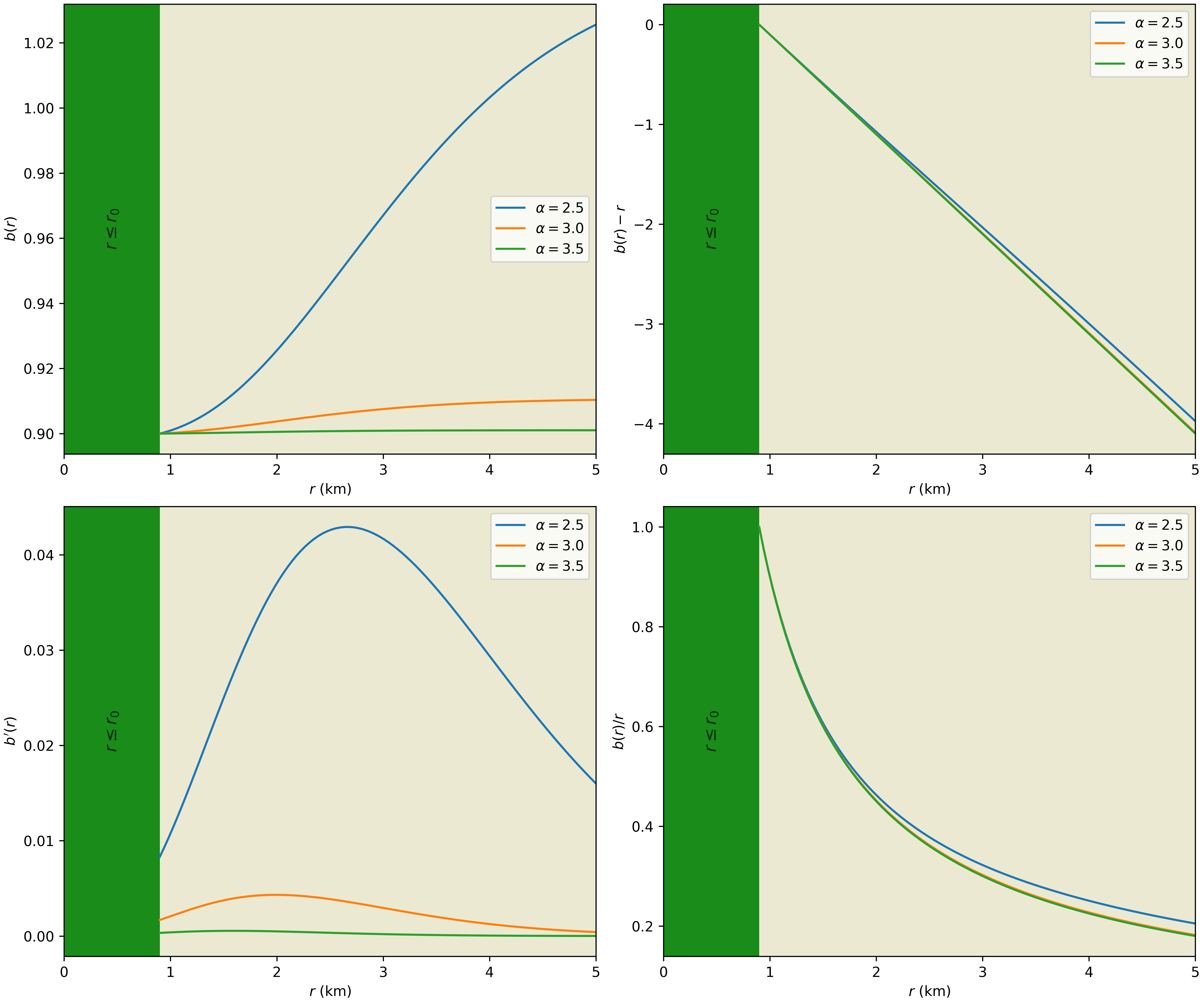}
    \caption{Nature of $b(r)$, throat condition $b(r) < r$, flaring-out condition $b'(r) < 1$, and asymptotic flatness $\lim_{r \to \infty} \frac{b(r)}{r} = 0$ for $r_{0}= 0.9$, $\rho_{0}=0.07$, and $\alpha=\{2.5,3.0,3.5\}$.}
    \label{shape2}
\end{figure}

The $\mathcal{E}^2$ selected as $\mathcal{E}^2=r^2$, we derive SF as 
\begin{align}\label{shape1.2}
 b(r)&=\frac{\rho_0^{\alpha -1} e^{\alpha  (-r)} }{\alpha^3} [ 
    2 (1-2 \alpha ) \rho_0 + (1-2 \alpha ) \alpha ^2 \rho_0 r^2  \nonumber\\
    &\quad 
    -\frac{\alpha ^4 e^r \left((\alpha -1)^4 r^4+4 (\alpha -1)^3 r^3+12 (\alpha -1)^2 r^2+24 (\alpha -1) r+24\right)}{(\alpha -1)^5}  \notag \\
    & \quad  + 2 (1-2 \alpha ) \alpha  \rho_0 r ]+D,
\end{align}
where D is obtained as
\begin{align}
    D&=r_0-\frac{\rho_0^{\alpha -1} e^{\alpha  (-r_0)} }{\alpha^3} [ 
    2 (1-2 \alpha ) \rho_0 + (1-2 \alpha ) \alpha ^2 \rho_0 r_0^2  \nonumber\\
    &\quad 
    -\frac{\alpha ^4 e^r_0 \left((\alpha -1)^4 r^4+4 (\alpha -1)^3 r_0^3+12 (\alpha -1)^2 r_0^2+24 (\alpha -1) r_0+24\right)}{(\alpha -1)^5}  \notag \\
    & \quad + 2 (1-2 \alpha ) \alpha  \rho_0 r_0 ].
\end{align}
The flaring-out condition is satisfied if the following conditions hold 
\begin{align}\label{15}
    (2\alpha-1) \rho_0^\alpha e^{{-\alpha r_0}}+ \alpha r_0^2 \rho_0^{\alpha-1}  e^{{-r_0 (\alpha-1)}}<\frac{1}{r_0^2}.
\end{align}
In the case of GR $\alpha=1$, hence
\begin{align}\label{6.9}
\rho_0<\left[\frac{1}{r_0^2}-r_0^2\right]e^{r_0}.
\end{align}
Here, we choose $r_0=0.9$, and hence $\rho_0<1.04$. We have fixed $\rho_0=0.07$ and by inequality (\ref{15}), we get value of $\alpha$ as $\alpha>0.84$.\\
\indent The graphical representation of SF with its different conditions given in the Fig. \ref{shape2}. The SF provided in Eq. (\ref{shape1.2}) has a similar graphical behavior to the previous one, looking after asymptotic flatness with $b(r)/r \rightarrow{0}$ as $r \rightarrow{\infty}$ and satisfies $b(r_{0})=0$ at $r_{0}=0.9$. Additionally, wormhole formation is ensured by upholding the requirement $b'(r)>0$. Even if the general trends are in agreement, little changes in parameters might result in somewhat different geometrical characteristics.

The radial pressure and tangential pressure are now
\begin{align}\label{6.10}
 \alpha p_r&=  (\alpha-1)\rho_0 e^{-r}+\alpha r^2 - \frac{\rho_0^{(1-\alpha)}e^{(\alpha-1)r}}{r^3} \Bigg(\frac{\rho_0^{\alpha -1} e^{\alpha  (-r)} }{\alpha^3} [ 
    2 (1-2 \alpha ) \rho_0 + (1-2 \alpha ) \alpha ^2 \rho_0 r^2 \Bigg.  \nonumber\\
    &\quad  \Bigg.
    -\frac{\alpha ^4 e^r \left((\alpha -1)^4 r^4+4 (\alpha -1)^3 r^3+12 (\alpha -1)^2 r^2+24 (\alpha -1) r+24\right)}{(\alpha -1)^5} \Bigg. \notag \\
    & \quad  \Bigg. + 2 (1-2 \alpha ) \alpha  \rho_0 r ]+r_0-\frac{\rho_0^{\alpha -1} e^{\alpha  (-r_0)} }{\alpha^3} [ 
    2 (1-2 \alpha ) \rho_0 + (1-2 \alpha ) \alpha ^2 \rho_0 r_0^2 \Bigg. \nonumber\\
    &\quad \Bigg.
    -\frac{\alpha ^4 e^r_0 \left((\alpha -1)^4 r^4+4 (\alpha -1)^3 r_0^3+12 (\alpha -1)^2 r_0^2+24 (\alpha -1) r_0+24\right)}{(\alpha -1)^5} \Bigg. \notag \\
    & \quad \Bigg. + 2 (1-2 \alpha ) \alpha  \rho_0 r_0 ]\Bigg),
    \end{align}
    \begin{align}\label{6.100}
 \alpha p_t&= (\alpha-1)\rho_0 e^{-r}-\alpha r^2+\rho_0^{(1-\alpha)}e^{(\alpha-1)r}\Bigg[\frac{1}{2r^3}\Bigg(\frac{\rho_0^{\alpha -1} e^{\alpha  (-r)} }{\alpha^3} [ 
    2 (1-2 \alpha ) \rho_0 + (1-2 \alpha ) \alpha ^2 \rho_0 r^2 \Bigg. \Bigg. \nonumber\\
    &\quad \Bigg. \Bigg.
    -\frac{\alpha ^4 e^r \left((\alpha -1)^4 r^4+4 (\alpha -1)^3 r^3+12 (\alpha -1)^2 r^2+24 (\alpha -1) r+24\right)}{(\alpha -1)^5} \Bigg. \Bigg. \notag \\
    & \quad \Bigg. \Bigg. + 2 (1-2 \alpha ) \alpha  \rho_0 r ]+r_0-\frac{\rho_0^{\alpha -1} e^{\alpha  (-r_0)} }{\alpha^3} [ 
    2 (1-2 \alpha ) \rho_0 + (1-2 \alpha ) \alpha ^2 \rho_0 r_0^2 \Bigg. \Bigg. \nonumber\\
    &\quad \Bigg. \Bigg.
    -\frac{\alpha ^4 e^r_0 \left((\alpha -1)^4 r^4+4 (\alpha -1)^3 r_0^3+12 (\alpha -1)^2 r_0^2+24 (\alpha -1) r_0+24\right)}{(\alpha -1)^5} \Bigg. \Bigg. \notag \\
    & \quad \Bigg. \Bigg.+ 2 (1-2 \alpha ) \alpha  \rho_0 r_0 ]\Bigg)- \frac{1}{2r^2}\Bigg(r^2  (2\alpha-1) \rho_0^\alpha e^{{-\alpha r}}+ \alpha r^2 \rho_0^{\alpha-1}  e^{{-r (\alpha-1)}} \Bigg) \Bigg].
\end{align}

\begin{figure}[ht]
    \centering
    \includegraphics[scale=0.4]{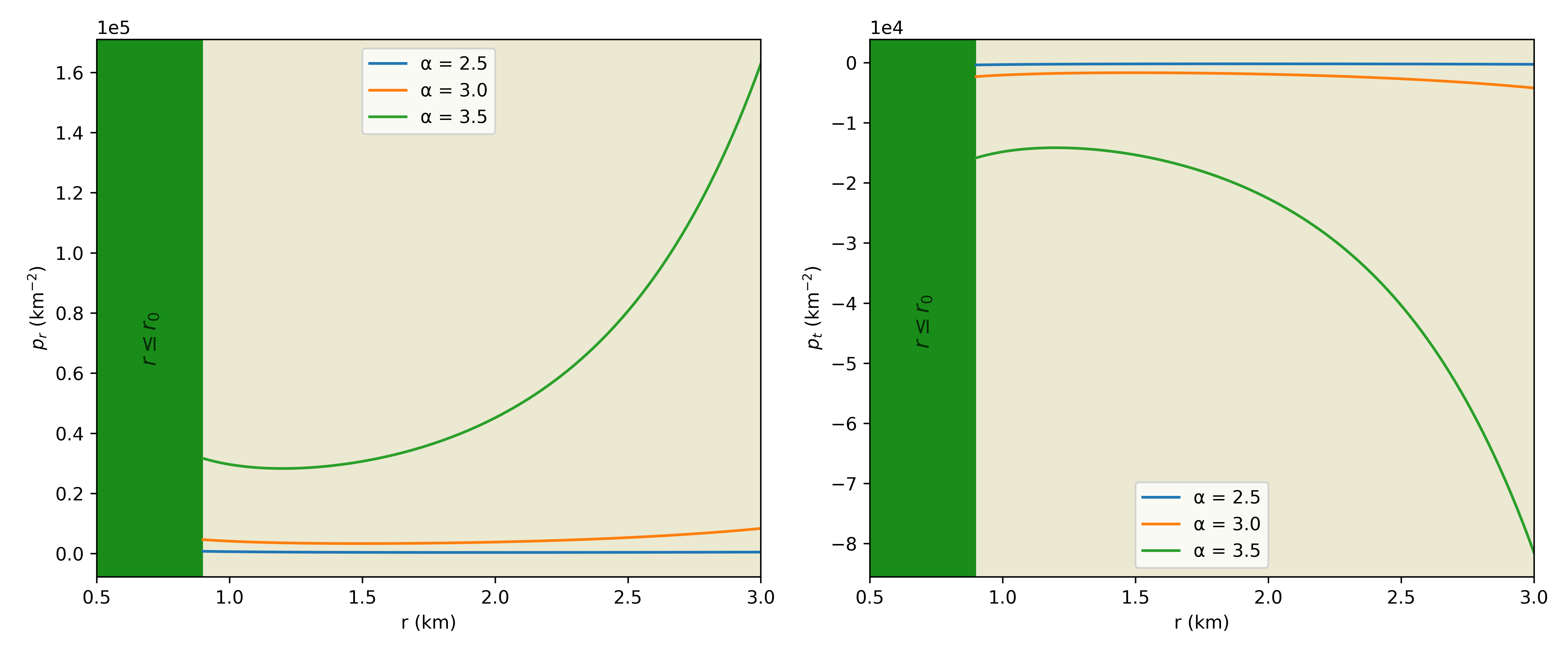}
  \caption{\textit{Left}: Radial pressure $p_r$ and \textit{Right}: Tangential pressure $p_t$ against the radius $r$ for the ES model. The graph shows different curve for the values of $\alpha$ for 2.5, 3.0, and 3.5., the density parameter $\rho_0=0.07$ and $r_0=0.9$ for the charge function $\mathcal{E}^2=r^2$.}
   
    \label{fig}
\end{figure}

For the SF in Eq. (\ref{shape1.2}), the radial pressure $p_{r}$ stays positive and evolves with the radial coordinate, whereas the tangential pressure $p_{t}$ is negative and falls with radius. This behavior differs from the prior SF, when $p_{r}$ was negative and $p_{t}$ was positive and increasing. The positive $p_{r}$ in this situation suggests a different force balance, which may affect the energy condition necessities. However, the anisotropic stress distribution supports the general stability of the wormhole structure.

 \begin{figure}
    \centering
    \begin{subfigure}[b]{0.45\textwidth}
        \centering
        \includegraphics[scale=0.5]{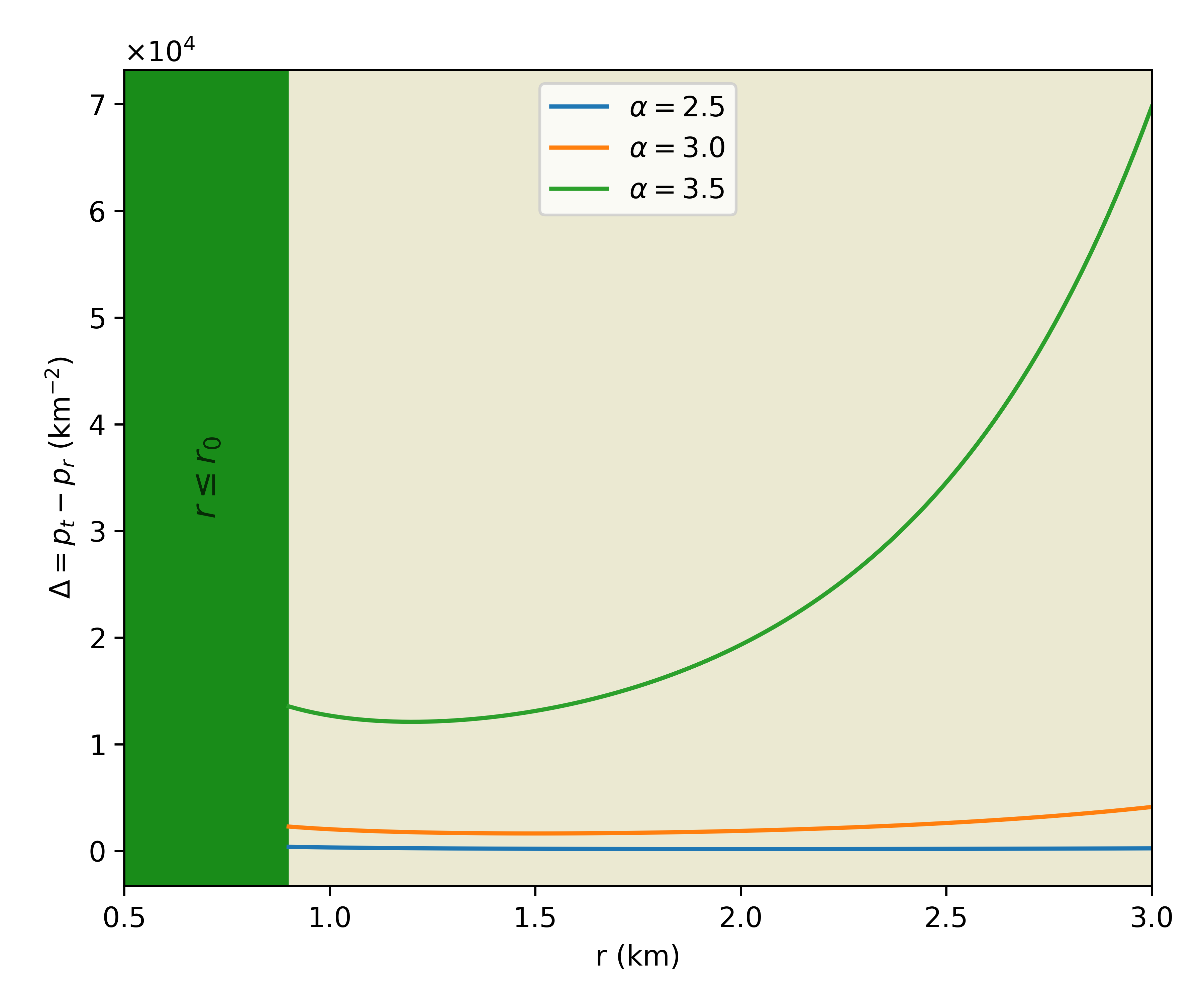}
        \caption{Anisotropy variation for the SF-$I$}
        \label{Delta-1}
    \end{subfigure}
    \hfill
    \begin{subfigure}[b]{0.45\textwidth}
        \centering
        \includegraphics[scale=0.5]{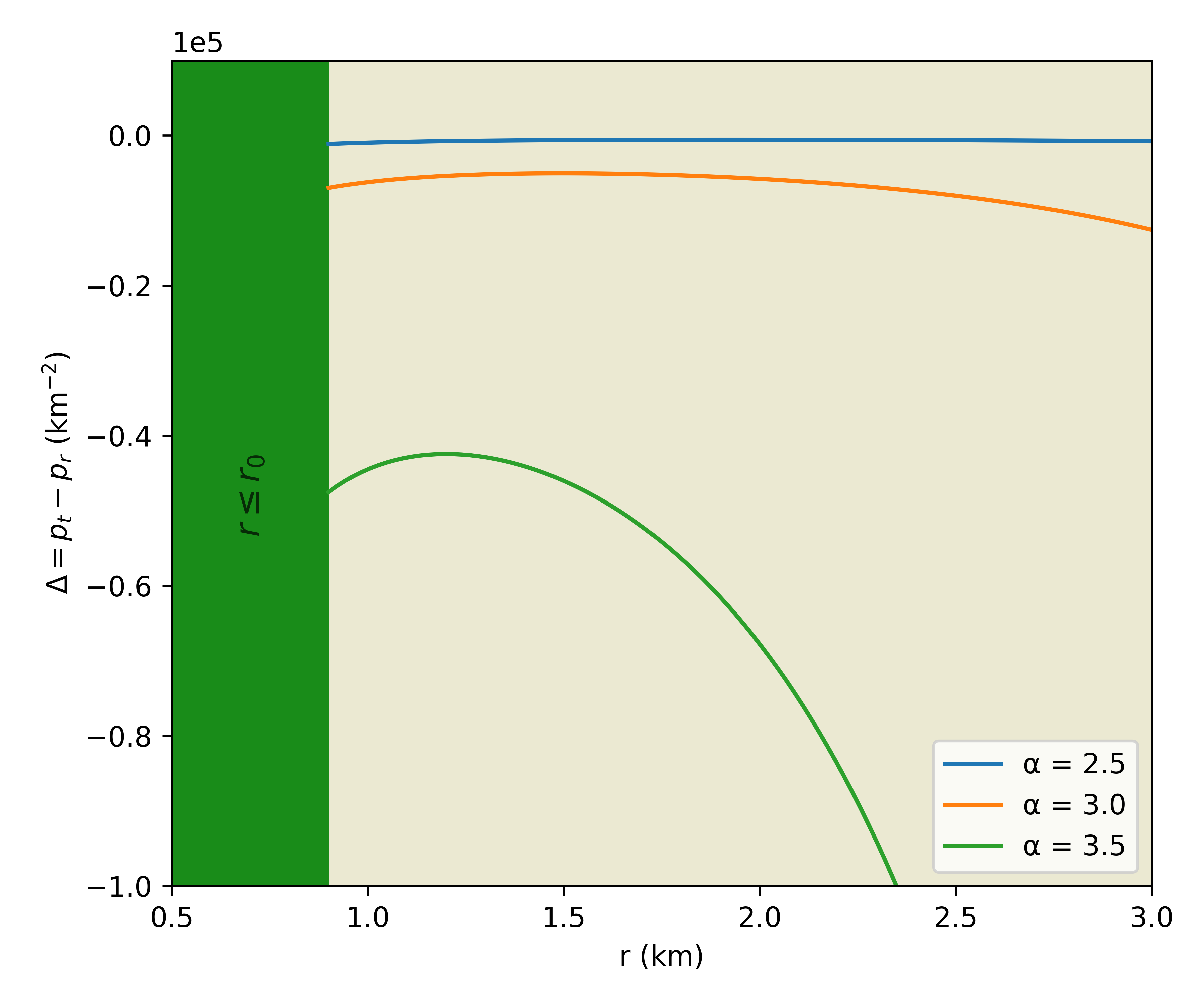}
        \caption{Anisotropy variation for the SF-$II$}
        \label{Delta-2}
    \end{subfigure}
    \caption{(a) Radial variation of anisotropy parameter $\Delta$ for the SF-$I$ and (b) Radial variation of anisotropy parameter $\Delta$ for the SF-$II$ with the parameters set to $A=0.5$, $r_0=0.9$, $\rho_0=0.07$, and different values of $\alpha$ 2.5, 3.0, 3.5.}
    \label{Delta-1,2}
\end{figure}
Figs. \ref{Delta-1,2} show the anisotropy profile $\Delta =p_{t}-p_{r}$ for both SFs. SF-1 has a negative $p_{r}$ and a positive $p_{t}$. This leads to a positive anisotropy parameter that increases with radial coordinate. The anisotropy ranges from $10^2$ to $10^4$  km$^{-2}$ , showing a significant tangential stress contribution.

On the other hand, for SF-2, where the $p_{r}$ is positive and the $p_{t}$ is negative, the anisotropy parameter is negative throughout the domain, as shown in Figure \ref{Delta-2}. The anisotropy ranges from $-10^2$ to $-10^4$  km$^{-2}$, showing an inverted stress distribution compared to SF-1.

An outward-directed anisotropy ($\Delta > 0$) is usually favorable for the existence of a traversable wormhole because it provides the required repulsive force to overcome gravity collapse while maintaining the wormhole throat. A typical feature of wormhole physics is the needed violation of the NEC, which can be made possible by a positive anisotropy, where $p_{t} > p_{r}$. On the other hand, a negative anisotropy ($\Delta <0$) indicates a radially dominating pressure, which could not obviously allow a traversable wormhole structure but might result in another type of stable exotic matter structure.

\subsection{\textbf{Case II}}
In this study, we explore the wormhole model by varying the values of the charge parameter while considering the EoS parameter with a fixed SF.\\
\indent Theoretically, the EoS parameter $\omega$ is an essential parameter, especially when examining gravitational theories, cosmology, and astrophysics \cite{carol}. To examine energy conditions for the wormhole, we have considered linear barotropic EoS. It establishes a relationship between the radial pressure $p_r$ and the energy density $\rho$ of given matter as
\begin{align}\label{eos}
p_r=\omega \rho.
\end{align}
\indent For a particular choice of SF taken as $b(r)=re^{1-\frac{r}{r_0}}$ with constant red-shift function and by considering EoS parameter given in (\ref{eos}), the field equations (\ref{efe4})-(\ref{efe6}) can be written as

\begin{eqnarray}\label{efe7}
    \rho=\frac{2\alpha \mathcal{E}^2}{[(\omega-1)\alpha+1]\frac{r}{r_0}-(\omega+1)\alpha},
\end{eqnarray}

\begin{eqnarray}\label{efe8}
   p_r=\frac{2\alpha \omega \mathcal{E}^2}{[(\omega-1)\alpha+1]\frac{r}{r_0}-(\omega+1)\alpha},
\end{eqnarray}

\begin{eqnarray}\label{efe9}
    &p_t=\frac{1}{\alpha}\left[\frac{e^{1-\frac{r}{r_0}}}{2rr_0}\left\{\frac{2\alpha \omega \mathcal{E}^2}{[(\omega-1)\alpha+1]\frac{r}{r_0}-(\omega+1)\alpha}\right\}^{1-\alpha}\right. \nonumber\\
  &\hspace{0.9cm}\left.+(\alpha-1)\frac{2\alpha \mathcal{E}^2}{[(\omega-1)\alpha+1]\frac{r}{r_0}-(\omega+1)\alpha}-\alpha \mathcal{E}^2 \right].
\end{eqnarray}

Eqs. (\ref{efe7}), (\ref{efe8}), and (\ref{efe9}) are the expressions for density, radial pressure, and tangential pressure, respectively.

\begin{figure}[ht]
    \centering
    \includegraphics[scale=0.4]{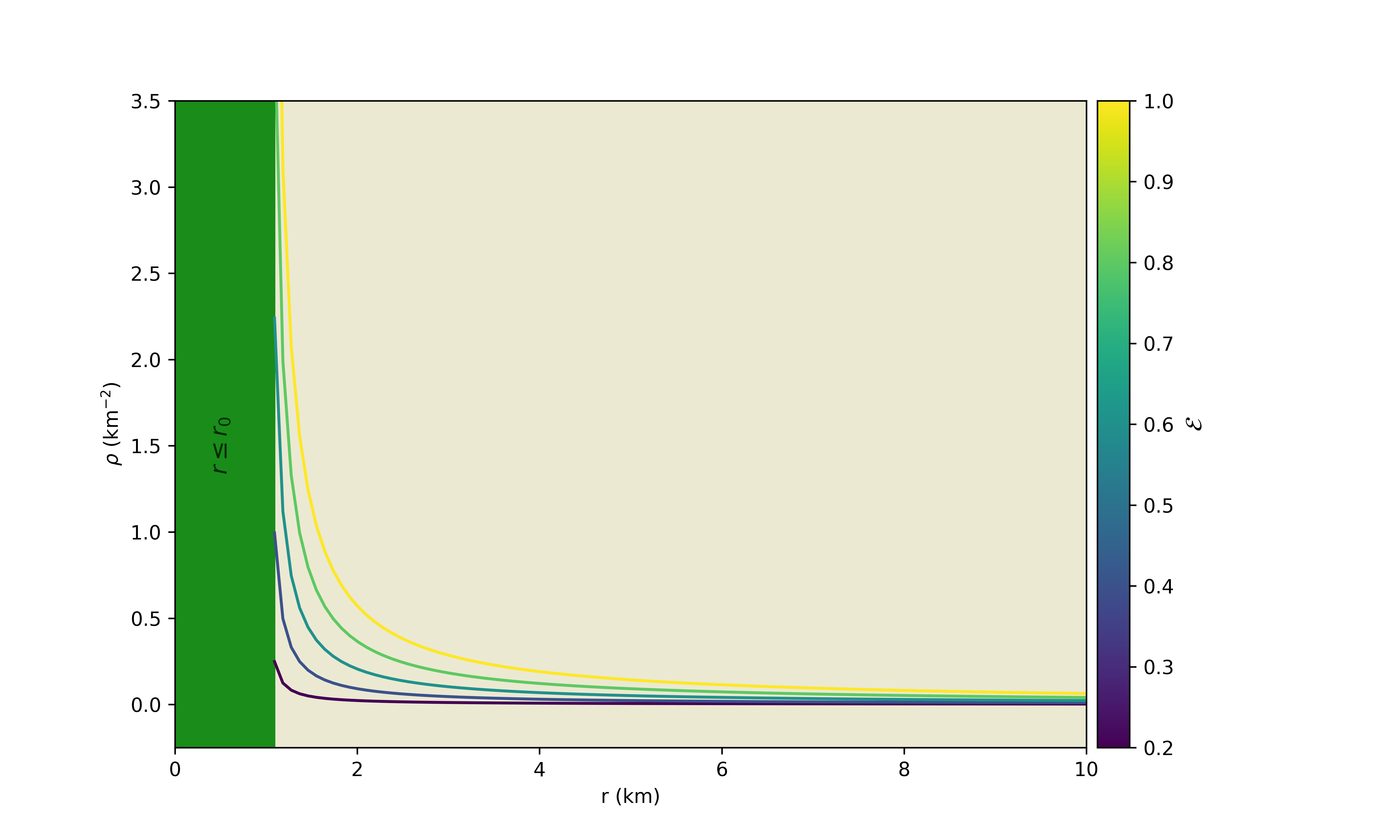}
    \caption{ Density \( \rho \) plotted against the radius \( r \) for a charged wormhole. The plot shows different curves for electric charge values $\mathcal{E} $ ranging from 0 to 1, with the model parameters set to \( \alpha = 0.5 \), EoS parameter \( \omega = 2.5 \), and throat radius \( r_0 = 1 \).}
    \label{fig_1}
\end{figure}

Fig. \ref{fig_1}, presents the density $\rho$ as a function of the radial coordinate $r$, computed via Eq.(\ref{efe7}) from the modified $f(R,\mathscr{L}_m)$ gravity. The model parameter is set at $\alpha = 0.5$, the EoS parameter at $\omega = 2.5$, and the electric charge $\mathcal{E}$ ranges from 0 to 1. The plot shows the density $\rho$ is greatest near the wormhole throat and progressively lowers as $r$ rises. In addition, we notice that greater magnitudes of $\mathcal{E}$ are correlated with increased densities around the throat. As the magnitude of the charge decreases the density curve accordingly lowers, emphasizing the apparent connection between electric charge and density distribution in the wormhole model.

\begin{figure}[htbp]
    \centering
    \begin{subfigure}[b]{0.45\textwidth}
        \centering
        \includegraphics[scale=0.3]{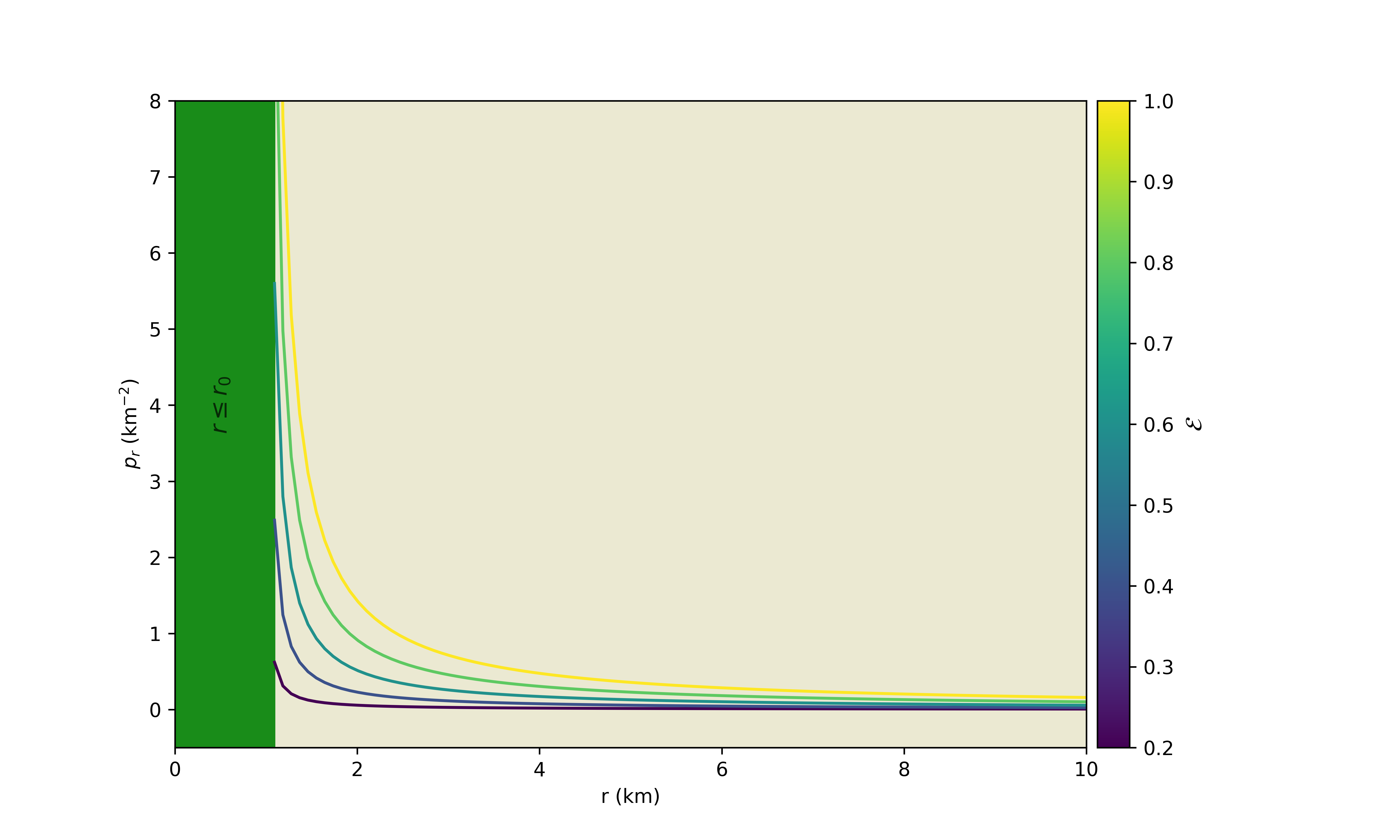}
        \caption{Radial pressure}
        \label{fig:radial-pressure}
    \end{subfigure}
    \hfill
    \begin{subfigure}[b]{0.45\textwidth}
        \centering
        \includegraphics[scale=0.3]{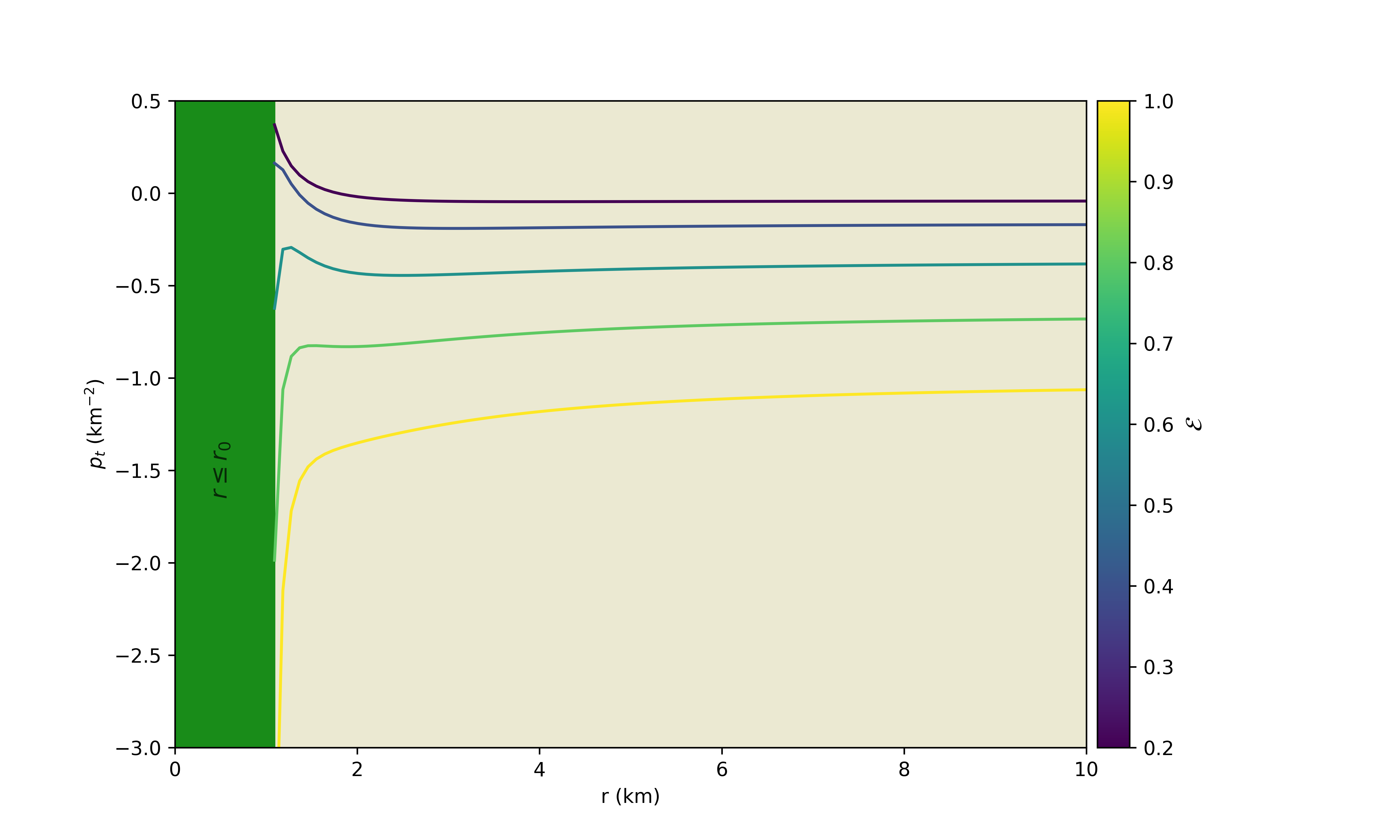}
        \caption{Tangential pressure}
        \label{fig:tangential-pressure}
    \end{subfigure}
    \caption{(a) Radial pressure \( p_r \) and (b) tangential pressure \( p_t \) plotted against the radius \( r \) for a charged wormhole. The graphs show different curves for electric charge values $\mathcal{E}$ ranging from 0 to 1, with the model parameters set to \( \alpha = 0.5 \), EoS parameter \( \omega = 2.5 \), and throat radius \( r_0 = 1 \). }
    \label{fig:pressure-comparison}
\end{figure}

In this study of wormholes adopting the framework of $f(R,\mathscr{L}_m)$ gravity, we check out the behavior of tangential and radial pressures as functions of the radial coordinate $r$, taking into account a particular range of charge values $\mathcal{E}$. The results show clear patterns in the pressure profiles based on the charge magnitude, as shown in Figs. (\ref{fig:radial-pressure}) and (\ref{fig:tangential-pressure}).

The profile for the $p_{r}$ shows positive values across the whole range that is taken into consideration, including the exterior region of the wormhole throat. However, for a high $\mathcal{E}$, the radial pressure reaches its highest value, demonstrating that stronger charge leads to increased radial pressure. The $p_{r}$ continuously falls with increasing radial distance from the throat, pointing to the anticipated reduction of the electromagnetic effect with distance. While in the under consideration range, the wormhole structure seems to be stable under the effects of the radial forces, as shown by the positive radial pressure throughout all $\mathcal{E}$ ranges.

The tangential pressure, $p_{t}$ on the other hand, exhibits more intricate behavior. Although it is lower in magnitude than the radial pressure $p_{r}$, the tangential pressure $p_{t}$ remain positive across the charge range of $\mathcal{E}= 0.1 \; \text{to} \; 0.35$.   This shows a decreased $p_{t}$ on the wormhole structure in the slight charge range. For the charge amount over $\mathcal{E}\approx 0.35$, $p_{t}$ becomes negative, suggesting a tension-like action rather than compression. The tendency toward negative values means the introduction of anisotropic pressure ($\Delta = p_{t}- p_{r} \neq 0$) by higher charges results in a decrease of $p_{t}$, which could suggest an instability in the direction of forces for larger charge magnitudes.

\begin{figure}
    \centering
    \includegraphics[scale=0.4]{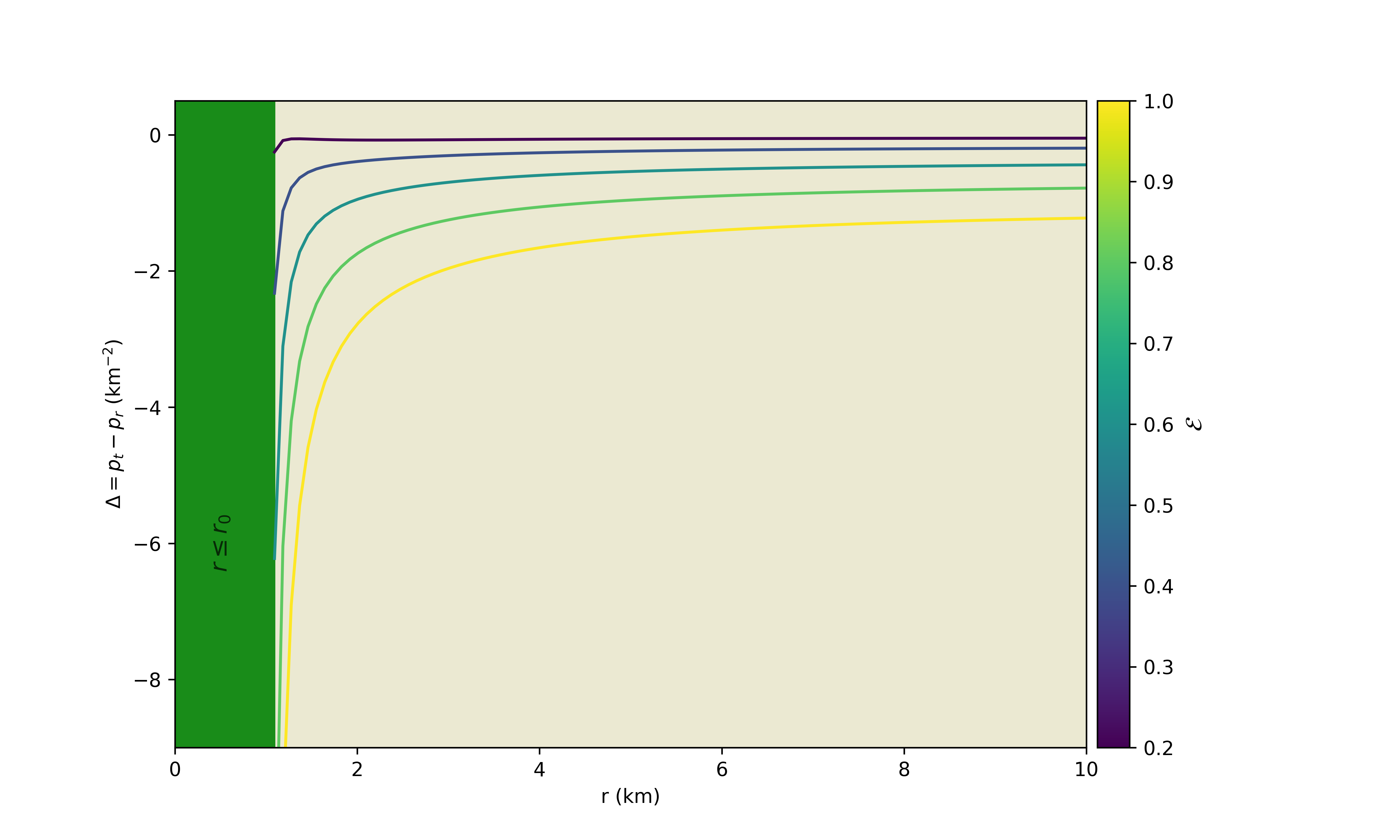}
    \caption{Radial variation of anisotropy parameter $\Delta$ for $ \alpha = 0.5 $,  $\omega = 2.5 $, and      $ r_{0} = 1 $. }
    \label{fig_3}
\end{figure}

Fig. \ref{fig_3} shows the anisotropy profile $\Delta$ as a function of  $r$, to show the interplay between pressures $p_{r}$ and $p_{t}$. The pressure $p_{r}$ is regularly larger than $p_{t}$ over the radial coordinate range. The structure provides a significant amount of $p_{r}$ in the region where $\Delta < 0$, signifying larger $p_{r}$ that maintain the wormhole's structure, which is  closest to the throat. This is crucial to keep the wormhole in its present shape to avoid any collapse driven on by the gravitational forces playing in a radial direction.

In a more plausible way, comparing this energy profile with the EoS of a neutron star-like compact object could provide a fair comparison between exotic matter and energy scale. Recently, many researchers investigated different kinds of exotic matter within the framework of highly gravitating compact objects; such exotic matter plays a crucial role in the anatomy of neutron stars and quark star-like objects.

 In Case II, the pressure-density profile shows that the matter density varies between around $0.1$ and $10$
km$^{-2}$, whereas the radial pressure varies within $1-16$
km$^{-2}$. However, in Case I, the pressure fluctuates within a much larger range of approximately $10^{3} - 10^{4}$ km$^{-2}$, while the density varies between about $10^{-3} - 10^{-2}$ km$^{-2}$. In compare to Case II, this shows that the pressure-density factors of different wormhole solutions can differ significantly, which at first may result in varying stability criteria and physical interpretations under strong gravitational fields. A few different EoS are compared in detail in Table \ref{tab:eos_comparison}, including those for wormhole configurations, geometrically obtained models, and neutron stars. In accordance with current understandings of nuclear matter, the first two EoS (SLy and SQM) are neutron stars. The density of these neutron$-$like compact stars ranges from around $10^{-28} -10^{-24}$ km $^{-2}$ ( in geometric units)\footnote{The density and pressure in geometric units \((\text{km}^{-2})\) can be converted to CGS units as follows: 
$$\rho_{\text{cgs}} = \rho_{\text{geom}} \frac{c^2}{G}, \quad p_{\text{cgs}} = p_{\text{geom}} \frac{c^2}{G},$$
where \(G = 6.67430 \times 10^{-8} \, \text{cm}^3 \, \text{g}^{-1} \, \text{s}^{-2}\) and \(c = 3 \times 10^{10} \, \text{cm/s}\).
:
$$
\rho_{\text{geom}} = \rho_{\text{cgs}} \frac{G}{c^2}, \quad p_{\text{geom}} = p_{\text{cgs}} \frac{G}{c^2}
$$}, whereas the pressure varies between $\sim 10^{-10}- 10^{-2}$ km$^{-2}$. The pressure-density relations are determined by spacetime geometry itself, as pointed out by the two geometrically obtained EoS (TRV and SNJR), which were calculated directly from Einstein's equations \cite{khunt_2021,khunt_2023}.

On the other hand, the wormhole EoS serves very differently in each case.  In Case II, the density typically ranges between $1$ km$^{-2}$ and $6$ km$^{-2}$, and the pressure varies between $1-16$ km$^{-2}$.  On the other hand, for Case I, the density typically ranges from $10^{-3}$ km$^{-2}$ to $10^{-2}$ km$^{-2}$, whereas the pressure varies between roughly $10^{3}$ km$^{-2}$ to $10^{4}$ km$^{-2}$. This is far more than that of typical neutron stars, demonstrating that wormholes as well as other unusual compact objects are impacted by gravitational fields that are far stronger. Wormholes are controlled by the curvature of spacetime itself, in comparison to neutron stars, which are made of nuclear matter.

 \begin{table}[h!]
    \centering
    \begin{tabular}{lccc}
    \multicolumn{4}{c}{\textbf{Compact stars}} \\
        \toprule
        \textbf{EoS} & \textbf{EoS type/model }  & \textbf{$\rho$ (km$^{-2}$)} & \textbf{$p$ (km$^{-2}$)} \\
       
        \midrule

        SLy \cite{SLy} & Nuclear matter&  $\sim 10^{-28}- 10^{-24}$ & $\sim 10^{-17}-10^{-3}$ \\
        SQM \cite{SQM} & Quark matter  &$\sim 10^{-26}-10^{-24}$  & $\sim10^{-4}-10^{-3}$ \\
        TRV \cite{khunt_2021,khunt_2023} & Geometrical & $\sim 10^{-23}-10^{-25}$ & $\sim10^{-8}-10^{-2}$ \\
        SNJR \cite{khunt_2021,khunt_2023} &  Geometrical & $\sim10^{-25}-10^{-24}$ & $\sim10^{-8}-10^{-5}$ \\

        \midrule
        \multicolumn{4}{c}{\textbf{Wormholes}} \\
        \midrule
         Solanki et al.\cite{solanki_2023}&$f(R,\mathscr{L}_m)$   & $\sim 1-8 $& $\sim 10^{-2}-10^{-1}$ \\
        
        Present study & $f(R,\mathscr{L}_m)$ gravity  (case I)  &  $\sim 10^{-3}-10^{-2}$ & $\sim 10^{3}-10^{4}$ \\

         & $f(R,\mathscr{L}_m)$ gravity (case II)&  $\sim 1-6$ & $\sim 1-16$ \\
        \bottomrule
    \end{tabular}
    \caption{Comparison of EoS for compact stars and wormholes, showing the type of EoS, density ($\rho$), and pressure ($p$) ranges. The data includes nuclear matter (SLy and SQM), geometrically derived models (TRV and SNJR), and wormhole configurations (Solanki et al.\cite{solanki_2023} and the present study). All values are expressed in geometric units (km$^{-2}$).
}
    \label{tab:eos_comparison}
\end{table}

This analogy emphasizes the important difference between exotic objects like wormholes and compact stars like neutron stars. The pressure-density profile of wormholes is determined by strong gravitational draws rather than material properties.
This observation offers significant insight concerning how to determine an appropriate geometric model for this sort of object. Since spacetime is a feature of geometry at its most fundamental, the most effective approach to understanding such strange objects is through gravity theory. We still are unaware of exactly what these objects are made of on their interiors. In addition, it further highlights the significance of further investigation to better understand such strong gravity objects and how different geometries influence their characteristics.

\section{Energy condition}
We study wormhole energy conditions determining if they follow conventional physical laws, such as the null energy condition, and to examine whether they are feasible to exist in the context of modified gravitational theories.

The energy-condition relations for traversable wormholes can be constructed based on their anisotropic distribution of matter \cite{mt,morris_1988}.
\begin{enumerate}
    \item Null Energy Condition: $\rho + p_{r} \geq 0$,\; $\rho +p_{t} \geq 0$,
    \item Weak Energy Condition: $\rho \geq 0$,\; $\rho+p_{r}\geq 0$, \;$\rho +p_{t}\geq 0$,
    \item  Dominant Energy Condition: $\rho \geq 0$, \; $\rho \pm p_{r} \geq 0$,\; $\rho \pm p_{t} \geq 0$,
    \item  Strong Energy Condition: $\rho+p_{r} \geq 0$,\; $\rho+p_{t}\geq 0$,\; $\rho+p_{r}+2p_{t}\geq 0$
\end{enumerate}

\begin{figure}[ht]
    \centering
    \includegraphics[scale=0.4]{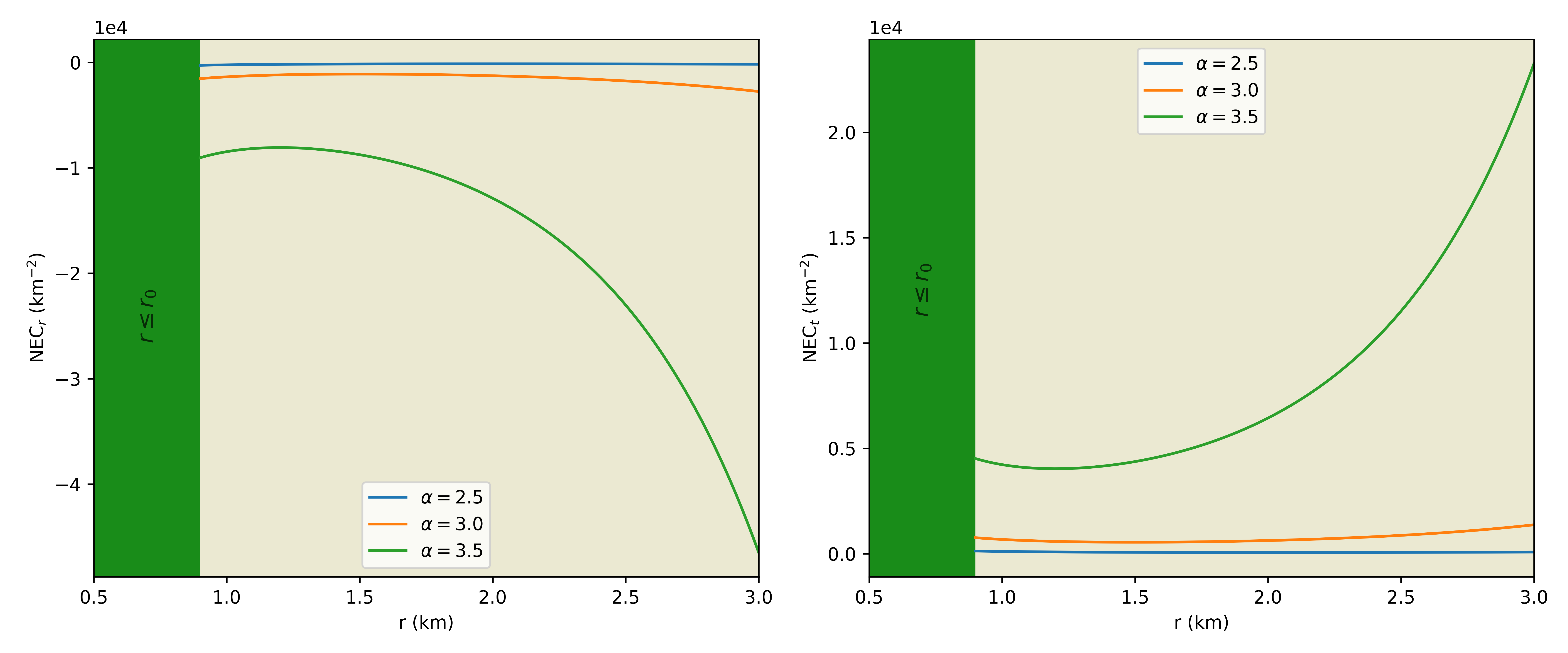}
    \caption{\textit{Left}: Radial NEC and \textit{Right}: Tangential NEC for SF-$I$}
    \label{nec1}
\end{figure}

\begin{figure}[ht]
    \centering
    \includegraphics[scale=0.4]{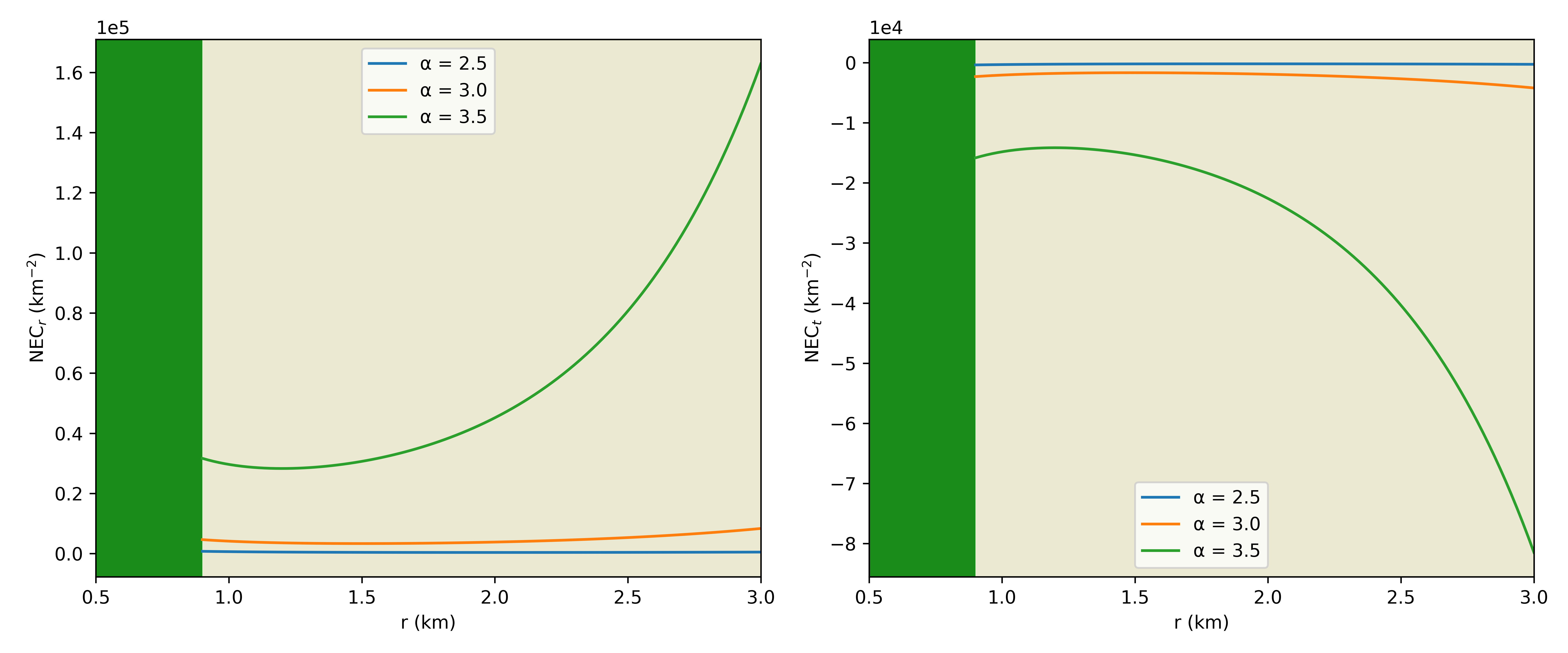}
      \caption{\textit{Left}: Radial NEC and \textit{Right}: Tangential NEC for SF-$II$}
    \label{nec2}
\end{figure}

\begin{figure}[htbp]
    \centering
    \begin{subfigure}[b]{0.45\textwidth}
        \centering
        \includegraphics[scale=0.3]{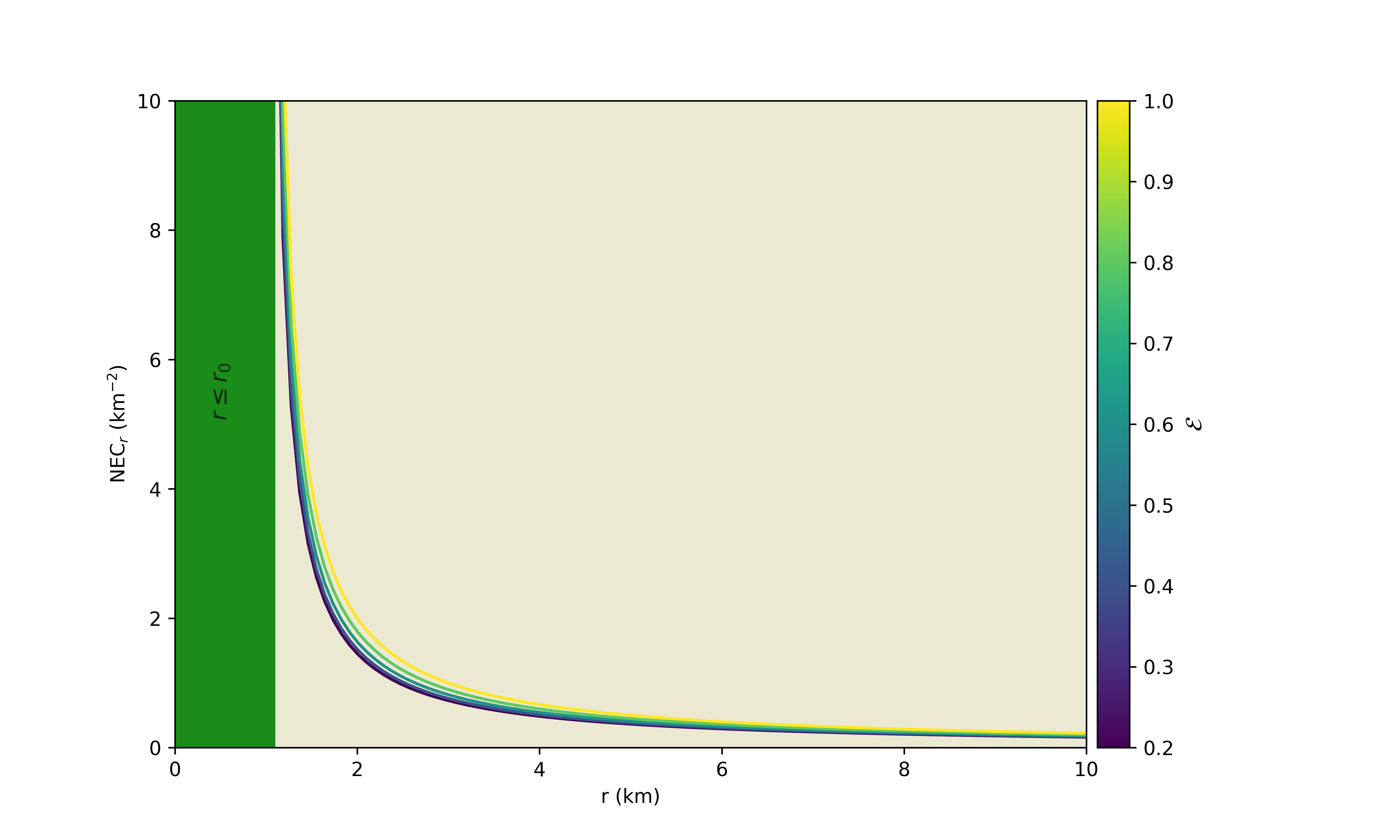}
        \caption{NEC$_{r}$}
        \label{fig:NECr}
    \end{subfigure}
    \hfill
    \begin{subfigure}[b]{0.45\textwidth}
        \centering
        \includegraphics[scale=0.3]{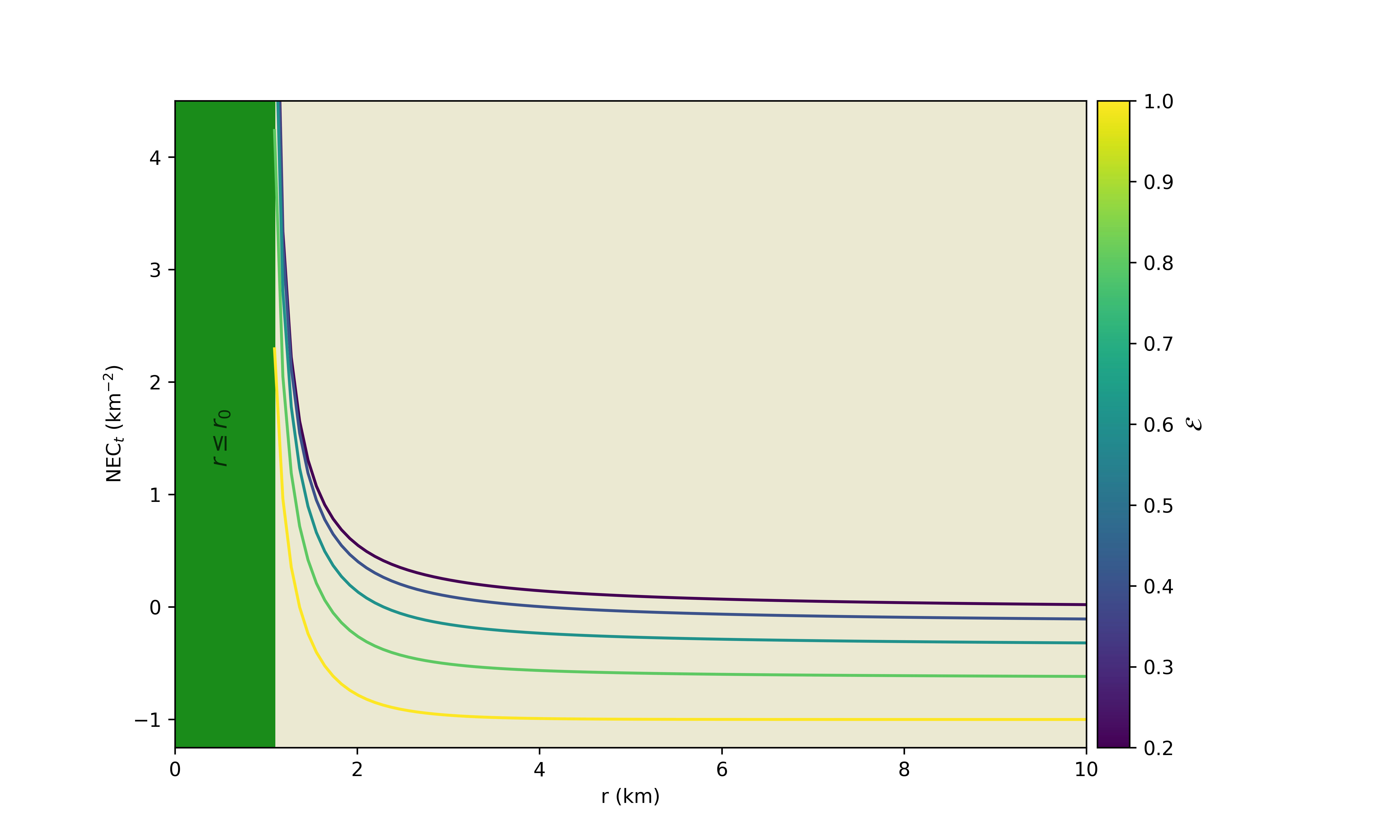}
        \caption{NEC$_{t}$}
        \label{fig:NECt}
    \end{subfigure}
    \caption{Null energy condition (NEC) for radial and tangential pressures for different values of $\mathcal{E}$. (a) NEC for radial pressure, which remains satisfied across the range. (b) NEC for tangential pressure, showing satisfaction within \( 0.1 \leq \epsilon \leq 0.6 \) and violation for $\mathcal{E} > 0.6$.}
    \label{fig:NEC}
\end{figure}

 The investigation on the energy conditions in $f(R,\mathscr{L}_m)$ gravity is elaborated in \cite{wang}. In the present study, we only present the NEC for the both cases, which is graphically represented in this section. \\
 \indent For each case, the NEC behaviour is shown for the tangential (NEC$_t$) and radial (NEC$_r$) components. For case-$I$, Fig. \ref{nec1} and \ref{nec2} show NEC$_r$ and NEC$_t$, respectively, for two different charge function choices. Fig. \ref{fig:NEC}, on the other hand, shows the NEC$_r$ and NEC$_t$ for case-$II$.
 
\indent We present the NEC for the ES density model, taking into consideration two distinct charge function choices, for both the NEC$_{r}$ and NEC$_{t}$ components.  The NEC is shown for the charge function $\mathcal{E}^2= A$ in Fig. \ref{nec1}, and for $\mathcal{E}^2 =r^2$ in Fig. \ref{nec2}.

The NEC$_{r}$ for $\mathcal{E}^2= A$ fails to fulfill the NEC in the radial direction as it violates the $\rho+p_{r}\geq 0 $ condition.  However, NEC$_{t}$  remains its positive contribution by following the NEC requirements.  As a result, NEC$_{t}$  exhibits a negative curve, which violates the NEC in the tangential direction, whereas NEC$_{r}$ shows a positive nature, satisfying the NEC for $\mathcal{E}^2 =r^2$.

As long as the $\mathcal{E}$ persists greater than a value of zero, the radial NEC$_{r}$ is satisfied. This is consistent from the $r_{0}$ to the maximum radius ($r_{max}$). In context, it indicates that the $p_{r}$ consistently supports the energy condition when a charge is involved.

On the other hand, NEC$_{t}$ shows a more restricted behavior. For $\mathcal{E}$ in the range $0.1 \leq \mathcal{E} \leq 0.6$, the NEC$_{t}$ is fulfilled, validating the energy condition within this interval. However, when the charge exceeds $\mathcal{E} > 0.6$, the NEC$_{t}$ is violated. Variations in tangential pressure cause this violation, leading to the formation of a throat-like environment that is crucial to maintaining the wormhole-like structure.

\section{Conclusion}
In the context of $f(R, \mathscr{L}_m)$ modified gravity, we investigated the possible existence of charged traversable wormholes. By taking energy momentum tensor as charged fluid, the field equations for $f(R, \mathscr{L}_m)$ gravity have been derived. For this study, the minimal coupling model $f(R, \mathscr{L}_m)=\frac{R}{2}+\mathscr{L}_m^\alpha$ has been considered. To simplify the expression of density and pressures, we have employed the EoS parameter along with specific shape and redshift functions.\\
\indent We have constructed two different shape functions based on the choices $\mathcal{E}^2=A$ and $\mathcal{E}^2=r^2$ using the ES model. For SF-I, Fig. \ref{nec1} shows that the NECr is not satisfied, while the NECt holds for all values of $\alpha$. On the other hand, for SF-II, Fig. \ref{nec2} indicates that NECt is not satisfied, whereas NECr holds for all values of $\alpha$. Therefore, for both shape functions (SF-I and SF-II), the NEC conditions are not fully satisfied.\\
\indent We have obtained the expression of energy conditions for the particular choice of SF $b(r)=re^{1-\frac{r}{r_0}}$. The existence of wormholes requires the fulfillment of the NEC. From Fig. (\ref{fig_1}), it can be observed that the energy density is positive throughout spacetime. Also, by (\ref{fig:NECr}), it can be seen that radial NEC is satisfied for all the values of $\mathcal{E} > 0$. However, as we can see from the Fig. (\ref{fig:NECt}), tangential NEC is valid only for the range $0.1 \leq \mathcal{E} \leq 0.6$. Therefore, NEC is valid only for $0.1 \leq \mathcal{E} \leq 0.6$ and hence for $\mathcal{E}>0.6$ results into existence of exotic matter.\\
\indent Moreover, the significance of incorporating modified gravity theories into our understanding of compact objects is shown by the significant differences in the energy profiles of charged wormholes and neutron stars. Neutron stars are well described by general relativity when matter is distributed in a standard way. However, charged wormholes in modified gravity models question these assumptions and suggest that exotic matter distributions may be required to sustain such structures. These different energy requirements show that general relativity can't fully explain non-standard events. These findings also demonstrate the potential of modified gravity theories to provide a more comprehensive framework for analyzing the physical properties of exotic objects such as wormholes.

\section*{Acknowledgements}
SVS is thankful to the CSIR, India for providing financial support under CSIR Senior Research Fellowship (09/157(0059)/2021-EMR-I), and also thankful to  IUCAA, Pune for the facilities and hospitality provided to him where the part of work was carried out.

\bibliographystyle{elsarticle-num} 
\bibliography{example}

\begin{thebibliography}{10}
\expandafter\ifx\csname url\endcsname\relax
  \def\url#1{\texttt{#1}}\fi
\expandafter\ifx\csname urlprefix\endcsname\relax\def\urlprefix{URL }\fi
\expandafter\ifx\csname href\endcsname\relax
  \def\href#1#2{#2} \def\path#1{#1}\fi

\bibitem{flm}
L.~Flamm, {Contributions to Einstein's theory of gravitation}, Hirzel, 1916.

\bibitem{ER}
A.~Einstein, N.~Rosen, {The Particle Problem in the General Theory of
  Relativity}, Phys. Rev. 48 (1935) 73--77.
\newblock \href {https://doi.org/10.1103/PhysRev.48.73}
  {\path{doi:10.1103/PhysRev.48.73}}.

\bibitem{geon1}
J.~A. Wheeler, Geons, Phys. Rev. 97 (1955) 511--536.
\newblock \href {https://doi.org/10.1103/PhysRev.97.511}
  {\path{doi:10.1103/PhysRev.97.511}}.

\bibitem{geon2}
C.~W. Misner, J.~A. Wheeler, {Classical physics as geometry}, Annals of Physics
  2~(6) (1957) 525--603.
\newblock \href {https://doi.org/https://doi.org/10.1016/0003-4916(57)90049-0}
  {\path{doi:https://doi.org/10.1016/0003-4916(57)90049-0}}.

\bibitem{mt}
M.~S. {M}orris, K.~S. {T}horne, {Wormholes in spacetime and their use for
  interstellar travel: A tool for teaching general relativity}, American
  Journal of Physics 56~(5) (1988) 395--412.
\newblock \href {https://doi.org/https://doi.org/10.1119/1.15620}
  {\path{doi:https://doi.org/10.1119/1.15620}}.

\bibitem{vis95}
M.~Visser, {Lorentzian Wormholes. From Einstein to Hawking}, Woodbury (1995).

\bibitem{Bron15}
K.~A. Bronnikov, A.~M. Galiakhmetov, {Wormholes without exotic matter in
  {E}instein–{C}artan theory}, Gravitation and Cosmology 21 (2015) 283--288.
\newblock \href {https://doi.org/https://doi.org/10.1134/S0202289315040027}
  {\path{doi:https://doi.org/10.1134/S0202289315040027}}.

\bibitem{soni24}
{Soni, Sagar V. and Khunt, A. C. and Hasmani, A. H.}, A study of
  {M}orris-{T}horne wormhole in {E}instein-{C}artan theory, International
  Journal of Geometric Methods in Modern Physics 21~(06) (2024) 2450115.
\newblock \href {https://doi.org/10.1142/S0219887824501159}
  {\path{doi:10.1142/S0219887824501159}}.

\bibitem{faruk21}
A.~Banerjee, A.~Pradhan, T.~Tangphati, F.~Rahaman, {Wormhole geometries in
  $\mathnormal{f(Q)}$ gravity and the energy conditions}, The European Physical
  Journal C 81 (2021) 1031.
\newblock \href {https://doi.org/10.1140/epjc/s10052-021-09854-7}
  {\path{doi:10.1140/epjc/s10052-021-09854-7}}.

\bibitem{rastgoo}
F.~Parsaei, S.~Rastgoo, P.~K. Sahoo, {Wormhole in $\mathnormal{f(Q)}$ gravity},
  The European Physical Journal Plus 137 (2022) 1083.
\newblock \href {https://doi.org/10.1140/epjp/s13360-022-03298-y}
  {\path{doi:10.1140/epjp/s13360-022-03298-y}}.

\bibitem{saibal24}
S.~Chaudhary, S.~Maurya, J.~Kumar, S.~Ray, {Stability analysis of wormhole
  solutions in $\mathnormal{f(Q)}$ gravity utilizing Karmarkar condition with
  radial dependent redshift function}, Astroparticle Physics 162 (2024) 103002.
\newblock \href
  {https://doi.org/https://doi.org/10.1016/j.astropartphys.2024.103002}
  {\path{doi:https://doi.org/10.1016/j.astropartphys.2024.103002}}.

\bibitem{symjan}
M.~Jan, A.~Ashraf, A.~Basit, A.~Caliskan, E.~Güdekli, {Traversable Wormhole in
  $\mathnormal{f(Q)}$ Gravity Using Conformal Symmetry}, Symmetry 15~(4)
  (2023).
\newblock \href {https://doi.org/10.3390/sym15040859}
  {\path{doi:10.3390/sym15040859}}.

\bibitem{Lobo2009}
F.~S.~N. Lobo, M.~A. Oliveira, {Wormhole geometries in $\mathnormal{f(R)}$
  modified theories of gravity}, Physical Review D 80~(10) (2009) 104012.
\newblock \href {https://doi.org/10.1103/PhysRevD.80.104012}
  {\path{doi:10.1103/PhysRevD.80.104012}}.

\bibitem{Bronnikov2007}
K.~A. Bronnikov, A.~A. Starobinsky, {No static spherically symmetric wormholes
  in scalar-tensor and $\mathnormal{f(R)}$ gravity}, JETP Letters 85~(1) (2007)
  1--5.
\newblock \href {https://doi.org/10.1134/S0021364007010018}
  {\path{doi:10.1134/S0021364007010018}}.

\bibitem{Pavlovic2015}
P.~Pavlovic, M.~Sossich, {Wormholes in viable $\mathnormal{f(R)}$ modified
  theories of gravity and Weak Energy Condition}, European Physical Journal C
  75~(3) (2015) 117.
\newblock \href {https://doi.org/10.1140/epjc/s10052-015-3347-3}
  {\path{doi:10.1140/epjc/s10052-015-3347-3}}.

\bibitem{Mazharimousavi2016}
S.~H. Mazharimousavi, M.~Halilsoy, {Wormhole solutions in $\mathnormal{f(R)}$
  gravity satisfying energy conditions}, Modern Physics Letters A 31~(16)
  (2016) 1650093.
\newblock \href {https://doi.org/10.1142/S0217732316500938}
  {\path{doi:10.1142/S0217732316500938}}.

\bibitem{Bahamonde2018}
S.~Bahamonde, U.~Camci, {Exact traversable wormhole solutions in
  $\mathnormal{f(R)}$ gravity by Noether symmetry approach}, Symmetry 10~(12)
  (2018) 774.
\newblock \href {https://doi.org/10.3390/sym10120774}
  {\path{doi:10.3390/sym10120774}}.

\bibitem{Moraes2017}
P.~H. R.~S. Moraes, P.~K. Sahoo, {Modeling wormholes in $\mathnormal{f(R, T)}$
  gravity}, Physical Review D 96~(4) (2017) 044038.
\newblock \href {https://doi.org/10.1103/PhysRevD.96.044038}
  {\path{doi:10.1103/PhysRevD.96.044038}}.

\bibitem{Godani2019}
N.~Godani, G.~C. Samanta, {Traversable wormholes and energy conditions with two
  different shape functions in $\mathnormal{f(R)}$ gravity}, International
  Journal of Modern Physics D 28~(2) (2019) 1950039.
\newblock \href {https://doi.org/10.1142/S0218271819500398}
  {\path{doi:10.1142/S0218271819500398}}.

\bibitem{Capozziello2021}
S.~Capozziello, A.~S. Ditta, E.~N. Saridakis, K.~Yin, {Traversable Wormholes
  with Vanishing Sound Speed in $\mathnormal{f(R)}$ Gravity}, European Physical
  Journal C 81~(2) (2021) 134.
\newblock \href {https://doi.org/10.1140/epjc/s10052-021-08996-y}
  {\path{doi:10.1140/epjc/s10052-021-08996-y}}.

\bibitem{ber2007}
O.~Bertolami, C.~G. B\"ohmer, T.~Harko, F.~S.~N. Lobo, {Extra force in
  $\mathnormal{f(R)}$ modified theories of gravity}, Phys. Rev. D 75 (2007)
  104016.
\newblock \href {https://doi.org/10.1103/PhysRevD.75.104016}
  {\path{doi:10.1103/PhysRevD.75.104016}}.

\bibitem{hlobo}
T.~Harko, F.~S.~N. Lobo, {$\mathnormal{f(R, L_m)}$ gravity}, The European
  Physical Journal C 70 (2010) 373--379.
\newblock \href
  {https://doi.org/https://doi.org/10.1140/epjc/s10052-010-1467-3}
  {\path{doi:https://doi.org/10.1140/epjc/s10052-010-1467-3}}.

\bibitem{lakhan1}
L.~V. Jaybhaye, S.~Bhattacharjee, P.~Sahoo, {Baryogenesis in $\mathnormal{f(R,
  L_m)}$ gravity}, Physics of the Dark Universe 40 (2023) 101223.
\newblock \href {https://doi.org/https://doi.org/10.1016/j.dark.2023.101223}
  {\path{doi:https://doi.org/10.1016/j.dark.2023.101223}}.

\bibitem{sahoo11}
N.~Kavya, V.~Venkatesha, G.~Mustafa, P.~Sahoo, S.~{Divya Rashmi}, {Static
  traversable wormhole solutions in $\mathnormal{f(R, \mathscr{L}_m)}$
  gravity}, Chinese Journal of Physics 84 (2023) 1--11.
\newblock \href {https://doi.org/https://doi.org/10.1016/j.cjph.2023.05.002}
  {\path{doi:https://doi.org/10.1016/j.cjph.2023.05.002}}.

\bibitem{sahoo12}
R.~Solanki, Z.~Hassan, P.~Sahoo, {Wormhole solutions in $\mathnormal{f(R,
  L_m)}$ gravity}, Chinese Journal of Physics 85 (2023) 74--88.
\newblock \href {https://doi.org/https://doi.org/10.1016/j.cjph.2023.06.005}
  {\path{doi:https://doi.org/10.1016/j.cjph.2023.06.005}}.

\bibitem{nasir}
T.~Naseer, M.~Sharif, A.~Fatima, S.~Manzoor, {Constructing traversable wormhole
  solutions in $\mathnormal{f(R, L_m)}$ theory}, Chinese Journal of Physics 86
  (2023) 350--360.
\newblock \href {https://doi.org/https://doi.org/10.1016/j.cjph.2023.10.032}
  {\path{doi:https://doi.org/10.1016/j.cjph.2023.10.032}}.

\bibitem{sahoo13}
L.~V. Jaybhaye, M.~Tayde, P.~K. Sahoo, {Wormhole solutions under the effect of
  dark matter in $\mathnormal{f(R, L_m)}$ gravity}, Communications in
  Theoretical Physics 76~(5) (2024) 055402.
\newblock \href {https://doi.org/10.1088/1572-9494/ad3746}
  {\path{doi:10.1088/1572-9494/ad3746}}.

\bibitem{mapari}
J.~Pawde, R.~Mapari, V.~Patil, et~al., {Anisotropic behavior of universe in
  $\mathnormal{f(R, L_m)}$ gravity with varying deceleration parameter},
  European Physical Journal C 84 (2024) 320.
\newblock \href {https://doi.org/10.1140/epjc/s10052-024-12646-4}
  {\path{doi:10.1140/epjc/s10052-024-12646-4}}.

\bibitem{mo_17}
P.~H. R.~S. Moraes, W.~de~Paula, R.~A.~C. Correa, {Charged wormholes in f(R,T)
  extended theory of gravity}, Int. J. Mod. Phys. D 28~(08) (2019) 1950098.
\newblock \href {https://doi.org/10.1142/S0218271819500986}
  {\path{doi:10.1142/S0218271819500986}}.

\bibitem{kimlee}
S.-W. Kim, H.~Lee, {Exact solutions of a charged wormhole}, Phys. Rev. D 63
  (2001) 064014.
\newblock \href {https://doi.org/10.1103/PhysRevD.63.064014}
  {\path{doi:10.1103/PhysRevD.63.064014}}.

\bibitem{eir2004}
E.~F. Eiroa, G.~E. Romero, {Linearized Stability of Charged Thin-Shell
  Wormholes}, General Relativity and Gravitation 36 (2004) 651--659.
\newblock \href {https://doi.org/10.1023/B:GERG.0000016916.79221.24}
  {\path{doi:10.1023/B:GERG.0000016916.79221.24}}.

\bibitem{j09}
J.~A. Gonz\'alez, F.~S. Guzm\'an, O.~Sarbach, {Instability of charged wormholes
  supported by a ghost scalar field}, Phys. Rev. D 80 (2009) 024023.
\newblock \href {https://doi.org/10.1103/PhysRevD.80.024023}
  {\path{doi:10.1103/PhysRevD.80.024023}}.

\bibitem{bro5}
K.~Bronnikov, S.~Grinek, {Conformal continuations and wormhole instability in
  scalar-tensor gravity}, Gravitation \& Cosmology 10 (2004).

\bibitem{j091}
J.~A. González, F.~S. Guzmán, O.~Sarbach, {Instability of wormholes supported
  by a ghost scalar field: I. Linear stability analysis}, Classical and Quantum
  Gravity 26~(1) (2008) 015010.
\newblock \href {https://doi.org/10.1088/0264-9381/26/1/015010}
  {\path{doi:10.1088/0264-9381/26/1/015010}}.

\bibitem{sharif14}
M.~Sharif, S.~Rani, {Charged Noncommutative Wormhole Solutions in
  $\mathnormal{f(T)}$ Gravity}, European Physical Journal Plus 129 (2014) 237.
\newblock \href {https://doi.org/10.1140/epjp/i2014-14237-5}
  {\path{doi:10.1140/epjp/i2014-14237-5}}.

\bibitem{mor}
P.~H. R.~S. Moraes, W.~de~Paula, R.~A.~C. Correa, {Charged wormholes in
  $\mathnormal{f(R, T)}$-extended theory of gravity}, International Journal of
  Modern Physics D 28~(08) (2019) 1950098.
\newblock \href {https://doi.org/10.1142/S0218271819500986}
  {\path{doi:10.1142/S0218271819500986}}.

\bibitem{h1}
T.~Harko, {Modified gravity with arbitrary coupling between matter and
  geometry}, Physics Letters B 669~(5) (2008) 376--379.
\newblock \href
  {https://doi.org/https://doi.org/10.1016/j.physletb.2008.10.007}
  {\path{doi:https://doi.org/10.1016/j.physletb.2008.10.007}}.

\bibitem{sahoo1}
L.~V. {Jaybhaye}, R.~{Solanki}, S.~{Mandal}, P.~K. {Sahoo}, {Cosmology in
  $\mathnormal{f(R, L_m)}$ gravity}, Physics Letters B 831 (2022) 137148.
\newblock \href {https://doi.org/10.1016/j.physletb.2022.137148}
  {\path{doi:10.1016/j.physletb.2022.137148}}.

\bibitem{hlobo2}
T.~Harko, F.~S. Lobo, {Generalized Curvature-Matter Couplings in Modified
  Gravity}, Galaxies 2~(3) (2014) 410--465.
\newblock \href {https://doi.org/10.3390/galaxies2030410}
  {\path{doi:10.3390/galaxies2030410}}.

\bibitem{l1}
N.~M. Garcia, F.~S.~N. Lobo, {Wormhole geometries supported by a nonminimal
  curvature-matter coupling}, Phys. Rev. D 82 (2010) 104018.
\newblock \href {https://doi.org/10.1103/PhysRevD.82.104018}
  {\path{doi:10.1103/PhysRevD.82.104018}}.

\bibitem{hlobo3}
T.~Harko, F.~S.~N. Lobo, J.~P. Mimoso, D.~Pav\'on, {Gravitational induced
  particle production through a nonminimal curvature\textendash{}matter
  coupling}, Eur. Phys. J. C 75 (2015) 386.
\newblock \href {https://doi.org/10.1140/epjc/s10052-015-3620-5}
  {\path{doi:10.1140/epjc/s10052-015-3620-5}}.

\bibitem{sofue13}
Y.~Sofue, {Rotation curve and mass distribution in the galactic center—from
  black hole to entire galaxy—}, Publications of the Astronomical Society of
  Japan 65~(6) (2013) 118.

\bibitem{carol}
S.~M. Carroll, {Spacetime and Geometry}: {An Introduction to General
  Relativity}, Cambridge University Press, 2019.
\newblock \href {https://doi.org/10.1017/9781108770385}
  {\path{doi:10.1017/9781108770385}}.

\bibitem{khunt_2021}
A.~Khunt, V.~Thomas, P.~Vinodkumar, {Distinct classes of compact stars based on
  geometrically deduced equations of state}, International Journal of Modern
  Physics D 30~(04) (2021) 2150029.
\newblock \href {https://doi.org/https://doi.org/10.1142/S0218271821500292}
  {\path{doi:https://doi.org/10.1142/S0218271821500292}}.

\bibitem{khunt_2023}
A.~Khunt, V.~Thomas, P.~Vinodkumar, {Relativistic stellar modeling with perfect
  fluid core and anisotropic envelope fluid}, Indian Journal of Physics 97~(12)
  (2023) 3379--3393.
\newblock \href {https://doi.org/https://doi.org/10.1007/s12648-023-02692-1}
  {\path{doi:https://doi.org/10.1007/s12648-023-02692-1}}.

\bibitem{SLy}
F.~Douchin, P.~Haensel, {A unified equation of state of dense matter and
  neutron star structure}, Astronomy \& Astrophysics 380~(1) (2001) 151--167.
\newblock \href {https://doi.org/https://doi.org/10.1051/0004-6361:20011402}
  {\path{doi:https://doi.org/10.1051/0004-6361:20011402}}.

\bibitem{SQM}
J.~Zdunik, {Strange stars-linear approximation of the EOS and maximum QPO
  frequency}, arXiv preprint astro-ph/0004375 (2000).

\bibitem{solanki_2023}
R.~Solanki, Z.~Hassan, P.~Sahoo, {Wormhole solutions in $\mathnormal{f(R,
  L_m)}$ gravity}, Chinese Journal of Physics 85 (2023) 74--88.
\newblock \href {https://doi.org/https://doi.org/10.1016/j.cjph.2023.06.005}
  {\path{doi:https://doi.org/10.1016/j.cjph.2023.06.005}}.

\bibitem{morris_1988}
M.~S. Morris, K.~S. Thorne, U.~Yurtsever, {Wormholes, time machines, and the
  weak energy condition}, Physical Review Letters 61~(13) (1988) 1446.
\newblock \href {https://doi.org/https://doi.org/10.1103/PhysRevLett.61.1446}
  {\path{doi:https://doi.org/10.1103/PhysRevLett.61.1446}}.

\bibitem{wang}
J.~Wang, K.~Liao, {Energy conditions in $\mathnormal{f(R, L_m)}$ gravity},
  Classical and Quantum Gravity 29~(21) (2012) 215016.
\newblock \href {https://doi.org/10.1088/0264-9381/29/21/215016}
  {\path{doi:10.1088/0264-9381/29/21/215016}}.

\end{thebibliography}
\end{document}